\documentclass{aa}  

\usepackage{graphicx}
\usepackage{txfonts}
\usepackage{orcidlink}
\usepackage{booktabs} 
\usepackage{placeins}
\usepackage{tabularx}

\begin{document}

   \title{Gamma-ray burst redshift estimation using machine learning and the associated web app}

   \subtitle{ }
    
   \author{A. Narendra\orcidlink{0000-0003-1265-2981}\fnmsep\thanks{First and second authors share the same contribution}
          \inst{1}\inst{2}
          \and
           M. G. Dainotti\orcidlink{0000-0003-4442-8546}\fnmsep\thanks{Corresponding author (maria.dainotti@nao.ac.jp)}\inst{3}\inst{4}\inst{5}\inst{6}\inst{7}
          \and
           M. Sarkar\orcidlink{0009-0007-0050-9762}\inst{9}
          \and
           A. {\L}. Lenart\orcidlink{0000-0003-1943-010X} \inst{1}
          \and
           M. Bogdan\orcidlink{0000-0002-0657-4342}\inst{12}\inst{13}
           \and
          A. Pollo\orcidlink{0000-0003-3358-0665}\inst{1}\inst{8}
          \and
          B. Zhang\orcidlink{0000-0002-9725-2524}\inst{11}
          \and
          A. Rabeda\orcidlink{0009-0002-9834-0611} \inst{1}
          \and         
          V. Petrosian\orcidlink{0000-0002-2670-8942}\inst{10}
          \and
          K. Iwasaki\orcidlink{0000-0002-2707-7548}\inst{3}\inst{4}\inst{14}
          }

   \institute{
        Astronomical Observatory of Jagiellonian University in Kraków, Orla 171, 30-244 Kraków
         \and
        Jagiellonian University, Doctoral School of Exact and Natural Sciences, Krakow, Poland
        \and
        Division of Science, National Astronomical Observatory of Japan, 2-21-1 Osawa, Mitaka, Tokyo 181-8588, Japan
        \and
        The Graduate University for Advanced Studies (SOKENDAI), Shonankokusaimura, Hayama, Miura District, Kanagawa 240-0115
        \and
        Space Science Institute, 4765 Walnut St Ste B, Boulder, CO 80301, USA
        \and
        Nevada Center for Astrophysics, University of Nevada, 4505 Maryland Parkway, Las Vegas, NV 89154, USA
        \and
        Bay Environmental Institute, P.O. Box 25 Moffett Field, CA, California
        \and
        National Center for Nuclear Physics (NCBJ), Warsaw
        \and
        Department of Physical Sciences, Indian Institute of Science Education and Research (IISER), Mohali, Punjab, India
        \and
        Department of Physics and Kavli Institute of Particle Astrophysics and Cosmology, Stanford University, Stanford, CA 94305, USA
        \and
        University of Nevada, Las Vegas, 4505 S. Maryland Pkwy, Las Vegas, NV 89154
        \and
        Department of Mathematics, University of Wroclaw, 50-384, Poland;
        \and
        Department of Statistics, Lund University, SE-221 00 Lund, Sweden;
        \and
        Center for Computational Astrophysics, National Astronomical Observatory of Japan, 2-21-1 Osawa, Mitaka, Tokyo 181-8588, Japan  
}
   \date{Received ; accepted }

% \abstract{}{}{}{}{} 
% 5 {} token are mandatory
 
  \abstract
   {Gamma-ray bursts (GRBs), which have been observed at redshifts as high as 9.4, could serve as valuable probes for investigating the distant Universe. 
   However, using them in this manner necessitates an increase in the number of GRBs with determined redshifts, as currently only 12\% of them have known redshifts due to observational biases.}
   {We aim to address the shortage of GRBs with measured redshifts to enable full realization of their potential as valuable cosmological probes.}
   {Following our previous approach, in this work we take a further step to overcome this issue by adding 30 more GRBs to our ensemble supervised machine learning training sample, representing an increase of 20\%, which will help us obtain more accurate pseudo-redshifts. In addition, we have built a freely accessible and user-friendly web application that infers the redshift of long GRBs (LGRBs) with plateau emission using our machine learning model. The web app is the first of its kind for such a study and will allow the community to obtain pseudo-redshifts by entering the GRB parameters into the app.}
   {Through our machine learning model, we successfully estimated redshifts for 276 LGRBs using X-ray afterglow parameters detected by the Neil Gehrels Swift Observatory and increased the sample of LGRBs with known redshifts by 110\%. We also performed Monte Carlo simulations to demonstrate the future applicability of this research.}
   {The results presented in this work will enable the community to increase the sample of GRBs with known pseudo-redshifts. This can help address many outstanding issues, such as GRB formation rate, luminosity function, and the true nature of low-luminosity GRBs, and it can enable the application of GRBs as standard candles.}

   \maketitle
%
%-------------------------------------------------------------------
\section{Introduction}

Gamma-ray bursts (GRBs) are the brightest and most powerful explosion events after the Big Bang and are observed across a wide range of redshifts, from 0.0085 \citep{Galma1998} to 8.2 and 9.4 \citep{Tanvir2008,Cucchiara2011}. 
This vast range of redshifts ($z$) makes GRBs ideal probes for studying the early Universe and tracking its evolution to gain insights into the composition, expansion rate, and other key aspects.
GRBs, generally detected in $\gamma$-rays, X-rays, and occasionally in optical wavelengths, are classified into two types based on their duration, which is characterized by the $T_{90}$ parameter. 
The parameter $T_{90}$ is the time span during which the GRB emits 90\% of its total observed fluence in $\gamma$-rays.
The GRBs with $T_{90}<2s$ are classified as short GRBs (SGRBs), and those with $T_{90}>2s$ are classified as long GRBs (LGRBs).
The LGRBs are produced by core collapse supernovae \citep{mazets1981catalog,kouveliotou1993identification,paczynski1998ApJ...494L..45P, Woosley2006ARA}, while SGRBs are produced by neutron star--neutron star or neutron star--black hole mergers
\citep{1992ApJ...395L..83N, 1992ApJ...392L...9D, 1992Natur.357..472U, 1994MNRAS.270..480T, 2007PhR...442..166N, 2017ApJ...846L...5G, 2017ApJ...848L..14G, 2017ApJ...848L..12A}.

In addition to these two main categories of GRBs, numerous other sub-classes have been identified in the literature.
These classes are based only on observational properties such as the $T_{90}$ duration as well as spectral and light curve (LC) properties, and they include classes such as SGRBs with extended emission (SEE; \cite{Norris2000, Norris2006, Levan2007, Norris2010}), where $T_{90}$ is greater than two seconds but with similar spectral properties to SGRBs, 
Very Long GRBs (VLGRBs, \cite{Levan2014}) and  Ultra Long GRBs (ULGRBs, \cite{stratta2013ultra,zhang2014long,Gendre2019}, where $T_{90}$ is greater than $500$ or
$1000$ seconds, respectively.
\cite{Zhang2007ApJ...655L..25Z} unified these various classes based on their GRB progenitors. 
According to their prescription, Type I GRBs comprise classes that are SGRBs, SEEs, and other GRBs similar to SGRBs and occur due to compact binary mergers \citep{2007PhR...442..166N, 2017ApJ...848L..12A}. 
Type II GRBs comprise LGRBs, VLGRBs, ULGRBs, and other similar classes with core-collapse supernova progenitors (the so-called collapsar events) \citep{woosley1993ApJ...405..273W, woosley1993}.
Irrespective of these efforts, an exhaustive classification of GRBs has not yet been found, and this is an active field of research.

{Phenomenologically, the GRB LC from across all the classes can be separated into two distinct phases.}
The prompt emission is the main emission, and it is characterized by a burst of high-energy photons spanning $\gamma$-rays to hard X-rays that sometimes extends into optical wavelengths \citep{Vestrand2005Natur,Beskin2010ApJ,2012MNRAS.421.1874G,2014Sci...343...38V}.
Following the prompt emission, the afterglow is a long-lasting, multiwavelength emission powered by the interaction of the GRB ejecta with the surrounding medium \citep{vanParadijs1997,costa1997,Piro1998}. 
It is detected across various wavelengths, such as X-rays, optical, and occasionally even radio.

A peculiar trait of GRBs, uncovered by the Neil Gehrels Swift observatory (hereafter Swift, \cite{Gehrels2004}), is the presence of a flat region in the LC, the so-called plateau emission, which is characterized by almost constant flux values
\citep{Nousek2006,Zhang2006,OBrien2006,Sakamoto2007,Liang2007,Dainotti2008,Zaninoni2013,Rowlinson2014}.
{The end of the plateau is considered to be a point in the LC's afterglow where the luminosity begins to decay with time as a power law.}
The plateau feature has been found in 42\% of X-ray afterglows \citep{Evans2009,Li2018b}.
{In our recent estimate, the percentage of GRBs with an observable plateaus is 38.5\%}.

Currently, the primary hurdle in the study of the LGRB population is the lack of a large GRB sample with known redshift.
Quick localization and spectral data are required to determine the redshift of GRBs.
Notably, Swift allows for this quick detection as well as multi-wavelength follow-up in $\gamma$-rays, X-rays, and ultraviolet/optical.
Swift has three telescopes: the Burst Alert Telescope (BAT; \cite{Burrows2005}), the X-ray Telescope (XRT; \cite{Barthelmy2005}), and the Ultraviolet/Optical Telescope (UVOT; \cite{Roming2005}). 
It uses XRT for precise localization, and with UVOT, it obtains spectra and sometimes redshifts.
Swift's rapid localization abilities have contributed significantly to exploring the high-$z$ Universe.

Although Swift triggers many observation programs, only 26\% (431) of its 1656 GRBs have known spectroscopic redshifts (as of 20 August 2024).
The Fermi Telescope has observed more than 3700 GRBs with its Large Area Telescope (LAT; \cite{Ajello2019}) and the Fermi Gamma-ray Burst Monitor (GBM; \cite{meegan2009}).
Currently, only 12\% of the GRBs detected by Swift and Fermi have spectroscopic redshift ($z_{obs}$), and approximately 90\% of them were observed by Swift.
Measuring redshift, especially high-$z$, is challenging due to limited telescope time and the difficulty of securing time for follow-up observations using large aperture telescopes.
Therefore, determining the $z$ of GRBs via alternative methods is crucial.

Correctly determining the GRB luminosity function (LF) is one of the main goals that can be achieved by increasing the number of GRBs with known redshifts. The LF provides the number of GRBs per unit luminosity and is vital for understanding the properties of the GRB population, such as their energy release and emission mechanisms.
Another principal objective is determining the cosmic GRB formation rate (GRBFR). The GRBFR measures the frequency of GRBs across cosmic time.

\cite{Petrosian2015,PetrosianDainotti2023arXiv230515081V} have found a discrepancy at low redshift ($0 < z < 1$) between the LGRB formation rate and the star formation rate (SFR).
This discrepancy, which has been observed by several research groups independently, remains an active area of investigation, and
addressing it requires a bigger sample of GRBs with known redshift, thus highlighting the importance of our study.

{The correct determination of the GRB rate for both LGRBs and SGRBs is hampered by the broad range of GRB physical features;
therefore, a precise distribution of these features and their evolution over cosmic time is necessary.
In \cite{Dainotti2024ApJ...967L..30D}, we investigated this issue with a sample of 179 GRBs with optical LCs, where we determined the luminosity and formation rate evolutions and the general shape of the LF after correcting for evolution.
We observed the average LGRB's rate at a low-$z$ ($z<1$) to be 10.46 Gpc$^{-3}$ $yr^{-1}$, which is roughly three times more than {\cite{Hopkins2006ApJ...651..142H} SFR (3.7 Gpc$^{-3}$ $yr^{-1}$).}
This LGRB rate is an actual rate where the normalization is in physical units, which contrasts with the arbitrary normalization used in most of the current literature.
}

To address the challenges outlined above, we propose a machine learning (ML)-based approach for estimating the redshift of GRBs.
This work is a follow-up of \cite{Dainotti2024GRBRedshift} and \cite{Dainotti2024ApJ...967L..30D}, and we utilize a sample that is 20\% larger than the sample in \cite{Dainotti2024GRBRedshift} to train our ML models. 
{In addition, we extend the generalization sample by 35\%}.
This sample is the set of GRBs without known redshifts, and we predict the redshift of 276 GRBs using our trained ML model.
Furthermore, we release a user-friendly web application that provides pseudo-redshifts for new GRBs built on the ML models described in this paper.

The paper is structured as follows.
In Sec. \ref{adv_ml}, we discuss the advantages of using an ML approach compared to using linear relations for the redshift estimation.
In Sec. \ref{datasample}, we describe our data set.
In Sec. \ref{methodology}, we describe the functionalities and various modules of our web app as well as
details of the ML models being used. We also describe the Monte Carlo (MC) simulation performed to understand the viability of our ML models in relation to future data.
Sec. \ref{results} shows the results obtained from our analysis and the usability of the web app.
Here, we also present the future applicability of our ML approach.
In Sec. \ref{discussion}, we discuss the relevance of our results. 
Finally, we summarize and conclude in Sec. \ref{conclusion}.

\section{Advantage of machine learning versus linear relationships}\label{adv_ml}
Expanding the GRB sample with redshift provides a significant benefit, notably the possibility of using them as standardized candles, {objects with known intrinsic luminosities, or with luminosities that can be derived via established relationships among GRB properties.}
The adoption of empirical relations between the distance-dependent and intrinsic properties of GRBs enables cosmological {studies in unexplored redshift ranges}.
One of the earliest efforts is the Dainotti Relation  \citep{Dainotti2008,Dainotti2011a,2015ApJ...800...31D,Dainotti2017}, a roughly inversely proportional relationship between the rest-frame time at the end of the plateau phase ($T_{a}/(1+z)$) and its corresponding X-ray luminosity ($L_{a}$).
Later, \cite{Dainotti2013} showed via the use of the Efron and Petrosian method (\cite{EffronPetrosian1992}), this relation is intrinsic and not due to selection biases {nor the redshift evolution}.
{The Dainotti Relation has also been observed in radio wavelengths \citep{dainotti2020b,Levine2022}.}
{It has also been extended in three dimensions in X-ray, optical \citep{Dainotti2022c}, and $\gamma$-rays \citep{2021ApJS..255...13D}, where the peak prompt luminosity ($L_{peak}$) has been added to the two-dimensional Dainotti relation \citep{Dainotti2016,dainotti2017b,Dainotti2020}. }

{The Dainotti relationships (both in two and three dimensions) were a subject of cosmological investigations, and }have been utilized as valuable tools \citep{2009MNRAS.400..775C, 2010MNRAS.408.1181C, 2013ApJ...774..157D, 2014ApJ...783..126P,Cao2021,CaoShulei2022,CaoShulei2022a,Dainotti2022b, Dainotti2023a, Bargiacchi2023, Dainotti2023b}.
{
The Dainotti relations have been shown to provide compatible results for $\Omega_M$ (the matter density parameter) under the assumption of flat $\Lambda$CDM model when used in conjunction with supernovae (SNe) Ia \citep{Dainotti2022b}.
This approach also has the advantage of expanding the distance ladder up to $z$ = 5, which is much higher than the most distant SNe Ia observed up to date ($z$ = 2.9) \citep{Pierel2024}.
{\cite{Wang2022ApJ...924...97W} further extended the Dainotti relation up to $z=5.91$. They also standardized the LGRB sample using a method similar to the Phillips Correlation \citep{Phillips1993ApJ...413L.105P}. This standardization helped \cite{Wang2022ApJ...924...97W} achieve a lower scatter on the rescaled luminosity, thus highlighting the importance of the plateau phase as a possible standard candle.}}

{
{Another notable effort in establishing a fundamental GRB relation is the so-called Combo relation \citep{Izzo2015A&A...582A.115I}, joining $L_a$, $T_{a}/(1+z)$, power-law decay of afterglow, and intrinsic spectral peak energy of prompt (E$_{p,i}$).
\cite{Izzo2015A&A...582A.115I} demonstrated, with a limited sample of 5 GRBs, a minimization of the dependency on SNe Ia for calibrating GRBs. They use a two-step calibration process for the Combo relation and measured $\Omega_M=0.29^{+0.23}_{-0.15}$ in $\Lambda$CDM.}
}

Recently, \cite{Shuang2022ApJ...924...69Y} analyzed samples of 174 X-ray LCs of GRBs with plateau and 104 GRBs with flares, overlapping by 51 sources. They discovered three tight correlations between plateau energy ($E_{plat}$), flare energy ($E_{flare}$) and isotropic prompt energy ($E_{\gamma,iso}$). The correlations were between $E_{plat}-E_{\gamma,iso}$, $E_{flare}-E_{\gamma,iso}$ and $(E_{plat}+E_{flare})-E_{\gamma,iso}$.
$E_{plat}-E_{\gamma, iso}$ correlation is similar to the $L_a-L_{peak}$ Dainotti correlation \citep{Dainotti2011}, thus independently confirming it.
\cite{Shuang2022ApJ...924...69Y} also conclude that X-ray plateaus and flares may originate in the same way
but, due to the effects of the surrounding medium, appear as different features in the afterglow LC.

The extension of the Dainotti relations in three dimensions has also been achieved by \cite{Xu2012A&A...538A.134X}, where they add the isotropic $\gamma$-ray energy to the two-dimensional relation, naming it the $L-T-E$ correlation. 
\cite{Zhao2019ApJ...883...97Z,Tang2019ApJS..245....1T} independently confirmed the existence of this correlation for 174 GRBs from Swift.
\cite{Tang2019ApJS..245....1T} also improved the best-fit relation to be $L_a \propto T_a^{-1.01}E^{0.84}$. 
This work was expanded upon further by \cite{XuFan2021ApJ...920..135X} and \cite{Deng2023ApJ...943..126D}.
\cite{XuFan2021ApJ...920..135X} showed the redshift evolution of the $L-T-E$ correlation and introduced a new correlation with the spectral peak energy labeled $L-T-E_p$ correlation.
\cite{Deng2023ApJ...943..126D} updated coefficients of the $L-T-E$ correlation of \cite{Tang2019ApJS..245....1T} to $L_a \propto T_a^{-0.99}E^{0.86}$
with 210 LGRBs with plateau and also derived pseudo-redshifts for 108 plateau GRBs.
However, their pseudo-redshifts are limited by intrinsic scatter and functional form of the $L-T-E$ correlation. 
Moreover, the precision of their approach is not tested against GRBs with known redshift.
In comparison, our ML-based approach tests the performance of our model on events with known redshifts.

{Previously, there have been several other efforts aiming to enhance the count of GRBs with known redshifts by exploring correlations between distance-independent parameters (such as peak flux and afterglow plateau duration) and distance-dependent GRB characteristics (such as prompt emission and peak luminosity) to estimate pseudo-redshifts for GRBs lacking observed values\citep{Reichart2001,Atteia2003,atteia2005,yonetoku2004,Dainotti2011}. 
Very few GRBs can reach the small uncertainty (5\%) \citep{Guiriec2016}. 
Specifically, when using a relation between the rest-frame $T_a$ and its luminosity, only 28\% of cases have uncertainties $\Delta z/z_{obs}<0.90$ \citep{Dainotti2011a}. 
If the redshift is extracted by employing the Amati correlation \citep{amati2002} between $E_{peak}$ (the peak in the $\nu F_{\nu}$ spectrum) and the energy emitted isotropically during the prompt emission,
the Pearson coefficient ($r$) of correlation between $z_{obs}$ and predicted redshift ($z_{pred}$) is $0.67$. 
\cite{amati2006} provided $z_{pred}$ for 17 GRBs, and these are accurate within a factor of 2.
}

Another often overlooked issue is the inherent inaccuracies of these pseudo-redshifts, which arise from their non-linear dependence on the luminosity distance ($D_L$).

{At low-$z$ since the $D_L(z)$ changes rapidly, even with parametric relations, we can infer the redshift, but as soon as $D_L(z)$ becomes flat around $z$ = 2, it is difficult to distinguish between the predictions among the high$z$ ($>2$), because a small variation of $D_L$ can still be reflected in a large variation of redshifts. 
In addition, $D_L(z)$ they have such a complex analytical dependence  $z$ that their relation cannot be determined analytically but only numerically.}

The major problem in the forward-fitting relationships is the dependence on cosmology. 
Indeed, one of the parameters of this relationship depends on the $D_L$, which itself depends on cosmological parameters.
{We also would like to stress that it is important to have a method that is not dependent on a given cosmological model; otherwise, we incur the so-called  "circularity problem". 
Indeed, it would be highly circular to use the pseudo-redshifts derived from the forward fitting method, which assumes a given cosmology to then derive cosmological parameters. 
Even if the goal of the pseudo-redshift is not to derive the cosmological parameters, we avoid being biased toward a particular or a given cosmological model for the analysis of the LF and density rate evolution here. 
Although, the error bars for the LF are not negligible, in the procedure of the selection biases detailed in \citep{Dainotti2024ApJ...967L..30D} if the cosmology had been inferred, we would have had a different number of associated sets (described in the appendix of that paper), because of the choice of a different limiting luminosity. Hence, this will also affect the density rate evolution.}

Thus, using any of these methods would induce a circularity problem, which must be avoided, especially when we tackle cosmology.
In addition, given the functional form of $\log(D_L)$, as it flattens at $z \gtrsim 3$, any small variation of $\log(D_L)$ can become a large variation in redshift determination.
Hence, the difficulty in determining the redshifts accurately with these forward-fitting methods.
Furthermore, based on relationships, the forward fitting methods mentioned do not account for the errors on the parameters, thus underestimating the realistic uncertainties on the redshifts.
Thus, we have followed a route via ML algorithms to obtain more accurate predictions and overcome these issues.

{Similar work has been done by \cite{Ukwatta2016,Aldowma2024MNRAS.529.2676A}. However, here we use the plateau features, whose correlations have less intrinsic scatter than prompt ones.
This can ensure more stable ML results.}

\section{The data sample}\label{datasample}

Due to the diverse nature of various GRB classes, it is essential to avoid mixing the characteristics of these classes in an ML analysis.
In a previous paper \citep{Dainotti2024GRBRedshift}, some members of our research team have pointed out that the removal of some additional classes, especially the SEE, does not significantly affect the prediction.
Thus, this study focuses only on LGRBs, the most numerous and homogeneous subset,
and all SGRBs were removed from our sample.
Our data comprises of LGRBs observed in $\gamma$-rays and X-rays by the BAT and XRT telescopes onboard the Swift.
They are found in the Third Swift-BAT GRB Catalogue \citep{Lien2016}, and for data from 2016 until {12th December 2023}, we use the \href{https://swift.gsfc.nasa.gov/archive/grb_table/}{NASA Swift GRB Search Tool}. 
This sample is an extension of the one presented in \cite{Dainotti2024GRBRedshift}, which itself was taken from \cite{Dainotti2020} and \cite{Srinivasaragavan2020}.
For training our models, we have removed the GRBs that have photometric redshifts; an example is the famous high-$z$ GRB090429B.
Each GRB has ten properties (features).
Four features belong to the prompt: 
the energy fluence over $T_{90}$ of the prompt emission in units of erg {cm}$^{-2}$ (\textit{Fluence}), 
$T_{90}$, 
the prompt peak photon flux in units of the number of photons cm$^{-2}$ s$^{-1}$ (\textit{Peak Flux}), and 
from the BAT Telescope, the prompt photon index of the photon energy distribution modeled with a power law (\textit{PhotonIndex}).
Six features belong to the X-ray plateau and afterglow: 
the time at the end of the plateau emission ($T_a$), 
its corresponding flux ($F_a$), 
the temporal power-law index after the end of the plateau emission ($\alpha$), 
the spectral index of the plateau ($\beta$) assuming a power law for the spectral energy distribution, 
the neutral hydrogen column density along the line of sight (\textit{NH}), 
the spectral index obtained as the time-averaged spectral fit from the Swift XRT Photon counting mode data ($\gamma$). 
The plateau emission feature is modeled using the so-called Willingale function (W07, \cite{Willingale2007}).

\begin{figure}
\centering
\includegraphics[width=0.49\textwidth]{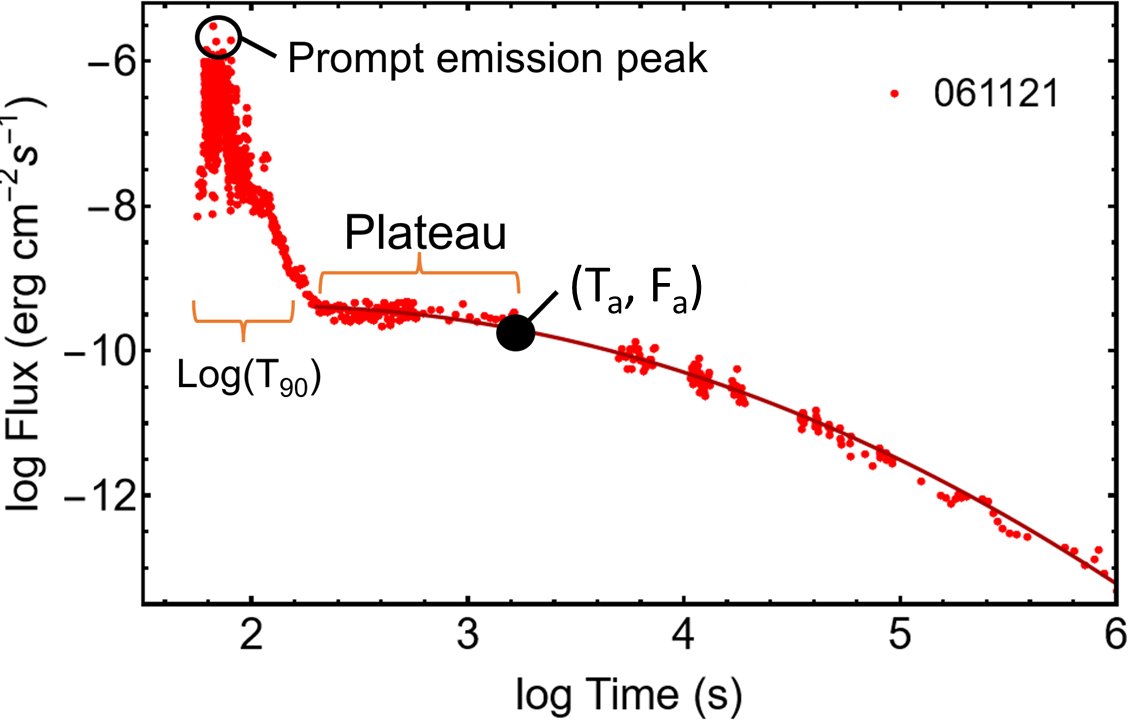}       %\vspace{-0.5em}
    \caption{\small
    \textit{Typical GRB LC. The X-ray LCs are fit with the W07 model, denoted with the darker red line. The end time of the plateau ($T_a$,$F_a$) and the prompt emission peak are shown as a black-filled dot and as an empty dot, respectively.}}
  \label{fig:1}
\end{figure}

Some of the parameters such as the $Peak flux$, $T_a$, $F_a$, and $T_{90}$, are shown the Fig. \ref{fig:1}, with $T_a$, $F_a$ representing the x and y coordinates of the black dot.
All of the parameters of the afterglow were obtained in 0.3-10 keV.

The initial data set contains 251 GRBs, including all classes. 
Compared with \cite{Dainotti2024GRBRedshift}, there are an additional 30 GRBs with redshift that are added to the sample, an increase of 13\%.
From this sample, we remove 13 SGRBs and are left with 238 LGRBs with all the features listed above, as well as our response variable, the $\log(z+1)$\footnote{We note that all the logarithms in this work are in base 10}. 
The response variable is the variable that we train our ML models to predict.
We call this set of 238 GRBs the training set.
This training set is 20\% larger than the previous training set of \cite{Dainotti2024GRBRedshift}.
For the generalization set, we use the 221 GRBs of \cite{Dainotti2024GRBRedshift} and add 78 GRBs (an increase of 35\%) taken from the Swift website.
The additional 78 GRBs of the generalization set were fit with the W07 model to obtain their $T_a$, $F_a$, $\alpha$, and $\beta$.

\begin{figure*}
    \centering
    \includegraphics[width=\textwidth]{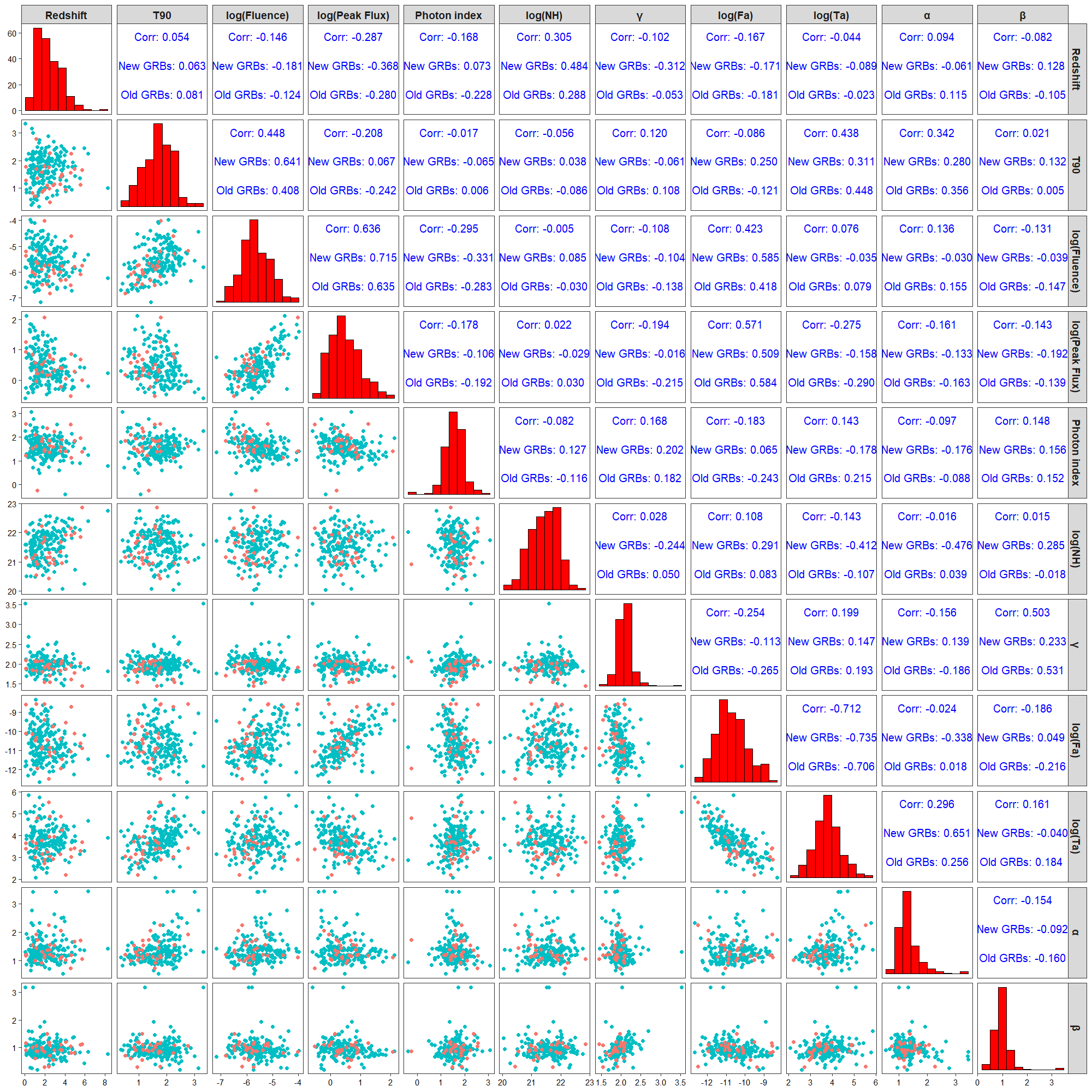}
    \caption{\textit{Scatter matrix plot showing the distribution of the features involved, with the diagonal showing the distributions.
    The 30 GRBs added to the training set are shown as orange dots, while those of the previous sample from \cite{Dainotti2024GRBRedshift} are represented by cyan dots. 
    The numbers in the upper triangle show the correlation between two respective features measured for new GRBs, the previous sample, and the combined sample.}
    }
    \label{fig:scattermatrix}
\end{figure*}

In Fig.\ref{fig:scattermatrix} we present the
scatter matrix plot between the training set from \cite{Dainotti2024GRBRedshift} (cyan) and the 30 new GRBs (orange) added in this analysis for comparing the distribution of the variables. 
We applied the Kolmogorov-Smirov test (KS test) {and the Anderson-Darling test (AD test)} for each variable between the two data sets. 

{
All the variables pass the two tests with a p-value $>$ 0.05,
indicating that they originate from the same parent population, 
except $T_{90}$ and $\gamma$, which obtain p-values $<$ 0.05 for both KS test and AD test, and $\log(F_a)$ which obtained a p-value $<$ 0.05 for AD test.
For further discussion see the Appendix and the Fig. \ref{fig:datacomparision}.
Here, we are presenting the comparison between the redshift distributions.
}

\begin{figure*}
    \centering
    \includegraphics[width=0.49\textwidth]{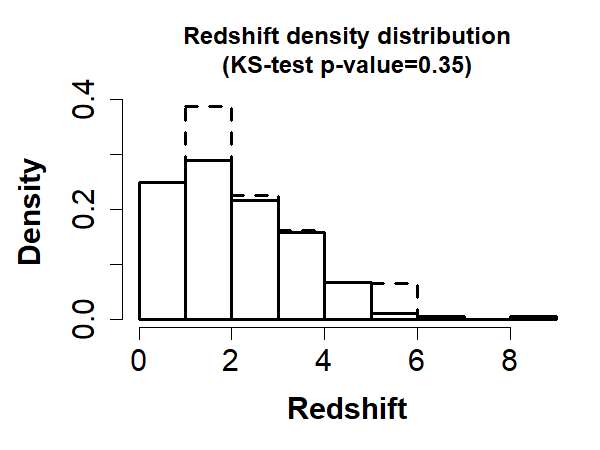}
    \includegraphics[width=0.49\textwidth]{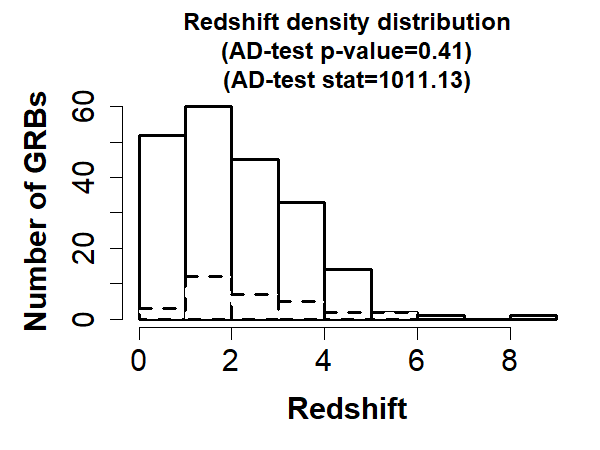}
    \caption{\textit{Left panel: Redshift distribution of the new sample (dashed lines) and the old sample (solid line) normalized to one. Right panel: Distribution of the new and the old samples when not normalized.}
    }
    \label{fig:redshift}
\end{figure*}

In the left panel of Fig. \ref{fig:redshift}, the redshift distribution is illustrated, with the new sample depicted by a dashed line and the old sample by a solid line, both normalized to one. 
The right panel of Fig. \ref{fig:redshift} presents the actual distribution with no normalization. 
{The redshift distributions pass the KS test with a p-value of 0.35 and the AD test with p-value of 0.41, as it can be seen from the right panel of Fig. \ref{fig:redshift} .}
{Thus, we cannot reject the hypothesis that the two sets of redshifts originate from the same parent population.}

\section{Methodology and web app}\label{methodology}

In this research, we expand the methodology implemented in \cite{Dainotti2024GRBRedshift}, and in addition, we make the webapp available to the community via an easy-to-use open-source web app. 
This web app is designed to let users input GRB parameters and quickly obtain pseudo-redshifts within hours of the GRB trigger.
This will allow the community to conduct follow-up observations of interesting bursts, such as high-$z$ events or low-$z$ low-luminosity GRB (LLGRB) events.
This web app is also highly modular, and users are free to train their own custom models using their data.

We here detail in each subsection the various methods implemented in our analysis and the associated functions of the web app.

\subsection{MICE}\label{sec:mice}

\begin{figure}[h!]
    \centering
    \includegraphics[width=0.49\textwidth]{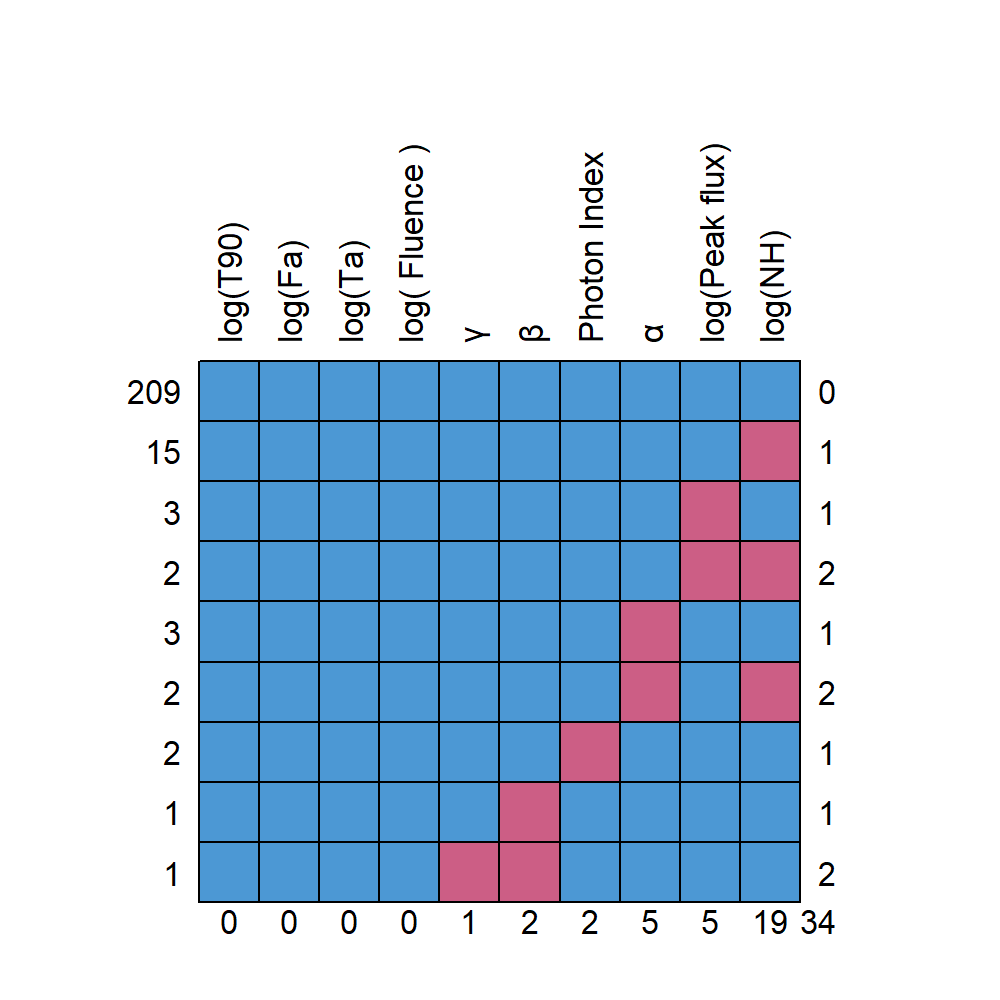}
    \caption{Distribution of missing data in our data set.
    The figure is divided into rows and columns. The top axis of the columns corresponds to the features in the data set, and the bottom axis corresponds to the total number of missing entries in that specific feature.
    Thus, for $\log(NH)$ there are 19 missing entries, for $\log(Peak flux)$ there are five missing entries, and so on.
    The left axis corresponds to the number of GRBs, and the right axis corresponds to the number of features with missing entries.
    The blue squares show the complete entries, and the red squares show which features are missing for a given number of GRBs.
    Thus, the first row indicates that there are 209 GRBs with no missing features, and the second row indicates that there are 15 GRBs with missing entries in only one feature, which is $\log(NH)$. 
    The third row indicates that there are three GRBs with missing entries in one feature, which is $\log(Peak flux)$, and the fourth row indicates that there are two GRBs with missing entries in two features simultaneously, which are $\log(Peak flux)$ and $\log(NH)$. 
    }
    \label{fig:mice}
\end{figure}

Similar to the sample in \cite{Dainotti2024GRBRedshift}, both our training and generalization sets have missing data.
Out of the 238 LGRBs composing our training set, 
29 GRBs have missing variables: 
15 GRBs have missing $\log(NH)$ entries, 
three GRBs have missing $\log(Peak flux)$ entries, 
two GRBs have missing entries in both $\log(Peak flux)$ and $\log(NH)$ simultaneously, 
three GRBs have missing entries in $\alpha$, 
two GRBs have missing entries in $\alpha$ and $\log(NH)$ simultaneously,
two GRBs have missing entries in $Photon Index$, 
one GRB has a missing entry in $\beta$ only,
and one GRB has missing entries in both $\gamma$ and $\beta$ simultaneously. (see Fig. \ref{fig:mice}).

In \cite{Dainotti2024GRBRedshift}, we removed the GRBs from the generalization set with missing entries. 
However, in this analysis, we impute them in a similar fashion to the training set. 
It should be noted that the imputation for the training set and the generalization set are conducted separately so that the training set does not influence the generalization set, thus preventing data bleeding.
All 299 GRBs in our generalization set are LGRBs, out of which 11 have missing data entries.
Two GRBs have missing entries in $\gamma$,
two have missing entries in $\log(NH)$,
two GRBs have missing entries in both $\gamma$ and $\log(NH)$ simultaneously,
two GRBs have missing entries in $\log(Peak flux)$ and $\log(T_{90})$
and three GRBs have simultaneous missing data in $PhotonIndex$, $\log(T_{90})$ and $\log(Peak flux)$.
In addition to the previously noted missing entries, we identify GRBs with $\log(NH) < 20$, $\alpha$, $\beta$, and $\gamma > 3$, and a $Photon Index < 0$ as outliers, as these values fall toward the extreme ends of their respective feature distributions.
Instead of removing these values from the analysis, we consider these as missing data and impute them alongside the previously mentioned missing data. 

To minimize data loss and enhance the size of both the training and generalization sets, we utilized the imputation technique known as multivariate imputation by chained equations (MICE; \cite{van2011mice,gibson2022}).
MICE leverages the data currently in use to infer the missing data. 
Here, for the data imputation to be reliable, MICE implicitly assumes that the data are missing at random. This means that the observed data can explain the missing data, and the probability of the data not being observed does not depend on the missing data’s value itself. Furthermore, MICE also assumes that the data have a multivariate distribution and thus considers the relations between features for imputing missing data.
Here, we stress that random imputation is appropriate, as the missing values of the variable are mainly due to the good and bad time intervals, orbital gaps, instrumental errors, and satellite follow-up repointing.

{The imputed data are sampled from the observed data, and the distribution is kept the same. 
We did this on purpose so that we would not have differences that could bias the analysis. 
In this study, we used the predictive mean matching method (PMM), specifically the mida-touch approach (see \cite{little2019statistical} for details). 
Using PMM ensures that the imputed data are sampled from the observed data.
This method fills in a feature's missing values with its mean and then iteratively refines these values by training a model on the complete data available. 
Each prediction is assigned a probability based on its proximity to the imputed value for the missing variable. The missing entry is then set by randomly choosing it from the observed values of the respective variable, weighted according to the previously defined probability. }
This process is carried out iteratively for each feature with missing entries until the values stabilize or a predetermined number of iterations is reached.
We perform this iteration 20 times to have a reliable imputation. 
With the application of MICE, we increase our training sample by 14\% and the generalization set by 4\%.

We also implemented MICE imputation in the web app so that any user can upload data with missing entries and obtain a complete data set with imputed values. 
The web app also shows the missing entries in the user's data, as shown in Fig. \ref{fig:mice} when the "Apply MICE" option is selected (see Fig. \ref{fig: App_SuperLearner1}).

\subsection{Machine learning methods}
 
 In the upcoming sections, we outline this study's machine learning methods and models.
 We discuss the feature selection methodology, the ML models used and the formula generation methodology.

\subsubsection{LASSO feature selection}\label{lasso}

\begin{figure}
    \centering
    \includegraphics[width=0.49\textwidth]{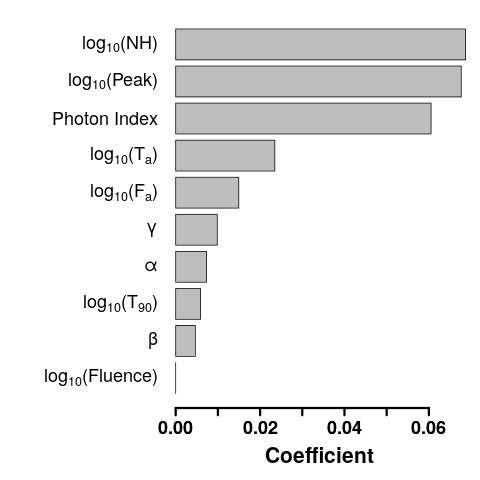}
    \caption{\textit{LASSO coefficients for all ten predictors. The top seven were selected for further analysis.}}
    \label{fig:lasso}
\end{figure}

The least absolute shrinkage and selection operator (LASSO) is a method for selecting the best features for a given response variable.
This method employs shrinkage for linear regression by constraining the $\ell^1$ norm (the sum of the magnitudes of all vectors in the given space) of the solution vector (in this case, $\log(z+1)$) to be less than or equal to a positive value called the tuning parameter ($\lambda$).
This penalization enables the model to choose a subset of features and eliminate the remaining by assigning a zero coefficient \citep{TibshiraniLasso}.
In this work, we apply the LASSO selection feature using the GLMNET function  \citep{hastie2017extended,tibshirani2012strong,Dainotti2021ApJ...920..118D}.
Then, we select the features' coefficients corresponding to the largest value of $\lambda$, with errors less than one standard deviation \citep{hastieTibs}.
The top seven features, $\log(NH),\log(Peakflux),PhotonIndex,\log(T_a),\log(F_a),\gamma$, and $\alpha$, (see Fig.\ref{fig:lasso}) are picked for further analysis.

\subsubsection{The machine learning models used}\label{testedmodels}

{
The ML models in general require a large amount of data to train. 
Here, even though we have the largest sample of LGRBs with plateau and redshift, this is still at least an order of magnitude less than the amount of data required to reliably train more complex ML models, such as neural networks.
}
Thus, we tested multiple different ML models to find the best set of models that can give accurate redshift predictions with the data set currently available.
The three chosen models performed the best during our analysis.
Furthermore, to test whether the performance of these models is maintained as the data set increases, we have performed a MC simulation as well (see Sec.\ref{MCMC}).
Thus, based on our analysis we are confident that these chosen models are best suited for the data at hand.

\begin{itemize}

\item Generalized additive model (GAM; \cite{hastie1990generalized}) is a semi-parametric method in which the response variable is connected to the predictors via the sum of either a parametric or non-parametric function or a combination of both. GAM provides the option for fitting smooth spline functions to the data that can enable non-linear fits. We explore the performance of GAM with and without the application of smoothing functions.

\item Generalized linear model (GLM; \cite{nelder1972generalized}) is an extension of the linear model that uses an iteratively weighted least squares method for the estimation of maximum likelihood to fit the relation between the response variable and the predictors.
GLM allows for non-linear distributions of the response variable through a link function. 
In our analysis, we use the default Gaussian link function. 

\item Random forest (RF) is a tree-based ML method that generates multiple independent regression trees \citep{Breiman2001MachL..45....5B,Miller2015ApJ...798..122M,Dainotti2021ApJ...920..118D}. The final results are an average of these regression trees.
RF has three hyperparameters: depth, number of trials, and number of trees. 
Hyperparameters are function parameters that determine a function's properties and dictate how well they fit a given set of data.
In our analysis, we use the default implementation of RF from the \textit{Caret} package \citep{caret}.

\end{itemize}

\subsubsection{Formula generation}\label{formulagen}

Following the methodology established by \cite{Dainotti2024GRBRedshift}, we aim to select the best formulas for GAM and GLM that can give us the most accurate predictions.

After performing LASSO-based feature selection to determine the set of the most predictive variables, we apply a method we call the formula generator. This method creates formulas based on combinations of these seven predictors and their squared terms. 
The sets of best formulas for GAM and GLM are determined independently. The process is described below.
One of the formulas that perform best with GLM is shown in Eq. \ref{equ} for demonstration purposes. The full list of formulas used in this analysis can be found on the web app and the text files provided in the associated GitHub repository.

To determine the best formulas for our data set, we divided the set of GRBs with $z_{obs}$ into a training set and a test set with an 80:20 split.
A total of 16510 formulas were created from combinations of the seven variables chosen by LASSO (see Sec. \ref{lasso}).
Each formula was tested by performing a ten-fold cross-validation (10fCV) with the training set 100 times.
The 10fCV is an ML method for assessing the performance of a model. 
It splits a given training set into ten parts (folds).
The model is trained on nine folds and predicts on the tenth fold. 
This step is repeated until the model has predicted for all ten folds, and then the results are assimilated, providing a measure of the model's performance.
We thus obtained a cross-validated Pearson correlation coefficient ($r$) and root mean square error (RMSE) values for each formula. 
Based on these scores, we select a subset of formulas that have achieved $r$ above 99.9\% quantile and RMSE score below 2\% quantile (see red points in the left panel of Fig. \ref{fig:formula_generation}). 
The exact cutoff for $r$ is 0.565, and for RMSE is 1.167.
This subset of formulas is used for obtaining predictions on the test set, which we had initially created (see right panel of Fig. \ref{fig:formula_generation}).
Finally, we selected the best formulas based on this test set's performance. Specifically, we picked three formulas that achieve the highest $r$, lowest median absolute deviation (MAD), and lowest RMSE:

\begin{figure*}
    \centering
    \includegraphics[width=0.49\linewidth]{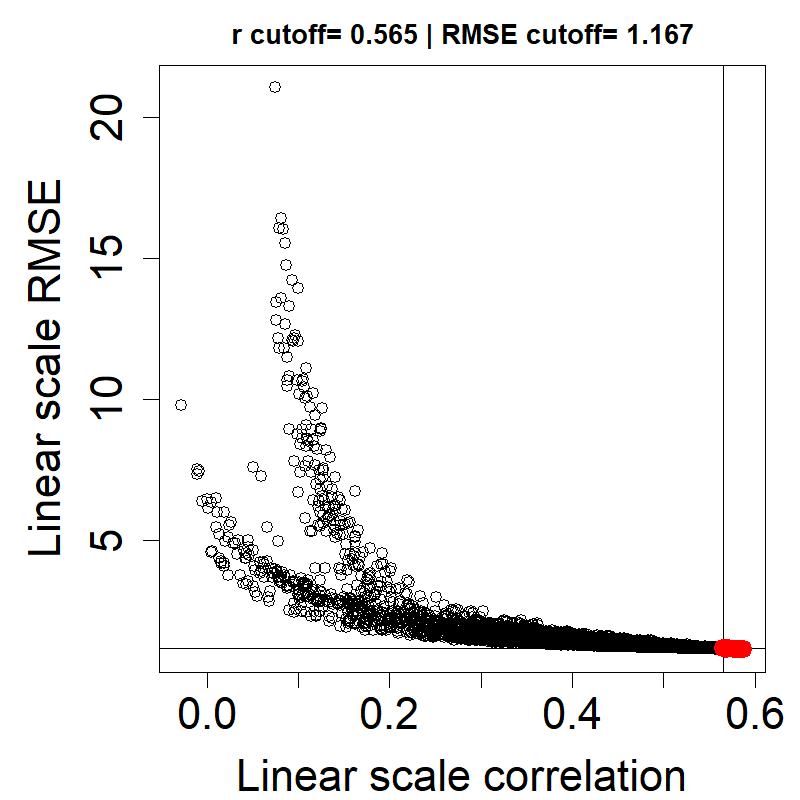}
    \includegraphics[width=0.49\linewidth]{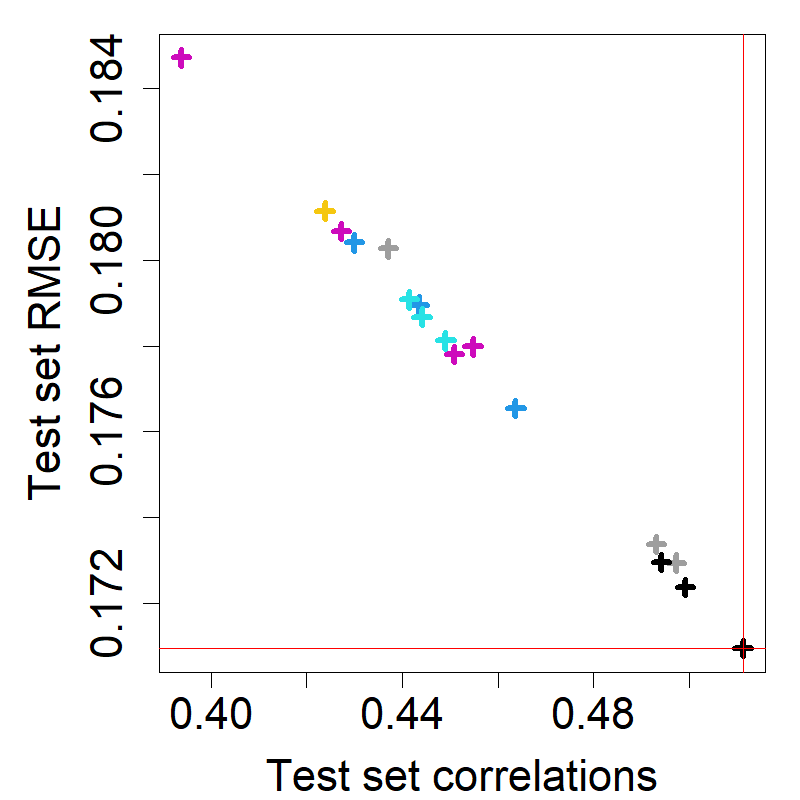}
    \caption{\textit{Left panel: Plot showing the RMSE and correlation scores for 16510 formulas. The red points show the formulas selected for analysis on the test set. 
    These formulas achieve correlation and RMSE scores above the cutoffs mentioned at the top of the figure.
    Right panel: Performance of the selected subset of formulas on the test set. Each plus symbol here represents one formula selected from the left panel (red points). The red vertical and horizontal lines point to the formula that obtained the lowest RMSE and highest correlation on the test set.}
    }
    \label{fig:formula_generation}
\end{figure*}

\begin{multline}
\begin{aligned} 
\log_{10}(z+1) = (\log_{10}(F_a)^2 + \log_{10}(F_a) + \log_{10}(PeakFlux))^2+\\
\log_{10}(NH) + PhotonIndex + \log_{10}(T_a) + \gamma + \alpha +\\
\log_{10}(NH)^2 + \log_{10}(PeakFlux)^2 + PhotonIndex^2 + \\
\log_{10}(T_a)^2 + \gamma^2 + \alpha^2.
\end{aligned}
\label{equ}
\end{multline}

However, during our analysis, we discovered that the formulas determined via the above-mentioned methodology are sensitive to the particular training set selected.
Thus, minimizing this strong dependence on the training set and selecting formulas that generalize to the full set of GRBs is important.
To accomplish this, we follow the steps mentioned above; however, we randomized the GRBs used in the training and test sets while maintaining the 80:20 split ratio.
We created 100 pairs of training and test sets and performed the subsequent steps for each pair. 
Furthermore, all the above-mentioned steps were repeated independently for GAM and GLM.

At the end of the computation, we counted which formulas were the best on the 100 different test sets and selected those that appeared the maximum number of times.
With this extension to the methodology, the formulas picked in this manner are generalized to the full GRB data set rather than a particular instance of training-test set combination.
Following this, we pick six formulas for GAM and four formulas for GLM.

In addition to testing the 16510 formulas with GAM and GLM, we also tested exclusively 148 formulas with GAM that have smoothing terms.
We note here that even though these 148 formulas with smoothing were included in the 16510 formula set, they were not chosen during the analysis. 
However, it is essential to test these because in our previous work, \citep{Dainotti2024GRBRedshift}, smoothing formulas worked better, as they can capture the non-linear relations between features. 
Thus, we decided to test them separately.
These 148 formulas include smoothing terms applied to features individually and in pairs. 
The methodology described above was applied to these 148 formulas, and we obtained two formulas that work best with our data set. These were included in our final ensemble.

We also tested formulas where three and four features were applied to the smooth function together. However, this did not improve our ML performance, and thus, we did not go forward with that set of formulas.

On the web app, we enable users to obtain the best formulas for their given data set.
However, the full methodology mentioned above has not been implemented due to computational limitations.
Performing 10fCV on 100 randomized training test sets for thousands of formulas requires extensive computational time.
We leveraged the supercomputing cluster at the Center for Computational Astrophysics at NAOJ, Tokyo, Japan and the PLGrid facility at Cyfronet, Krakow, Poland for our analysis and results.
This, however, is beyond the scope of a web application at this stage.
The formulas provided by the web app are determined on a single randomized training test set, similar to \cite{Dainotti2024GRBRedshift}.
Furthermore, the formulas are generated using only the variables, not their squared terms.
However, formulas with smoothing terms are included within this.

\begin{figure}
    \centering
    \fbox{\includegraphics[width=0.49\textwidth]{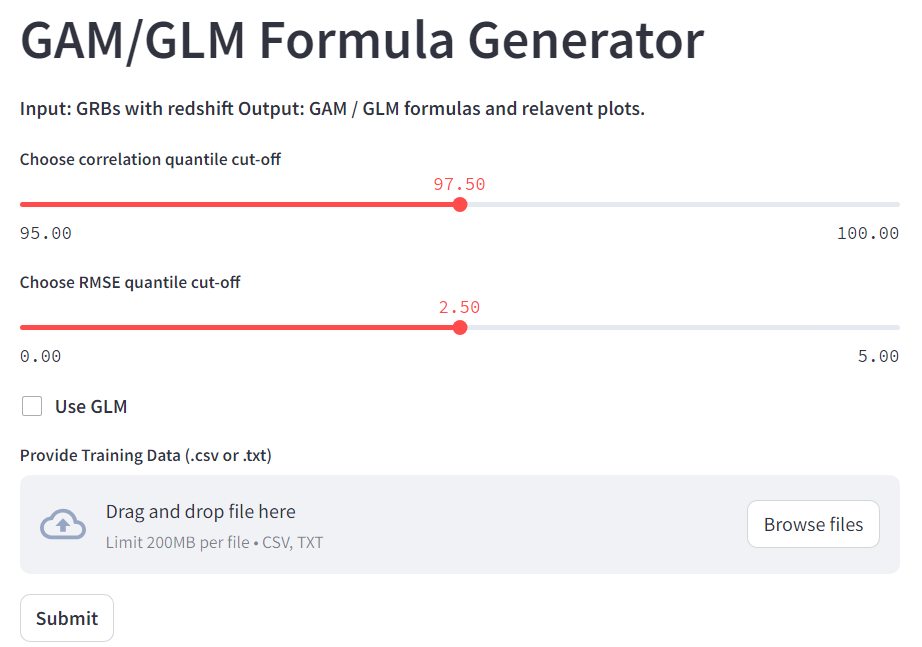}}
    \fbox{\includegraphics[width=0.49\textwidth]{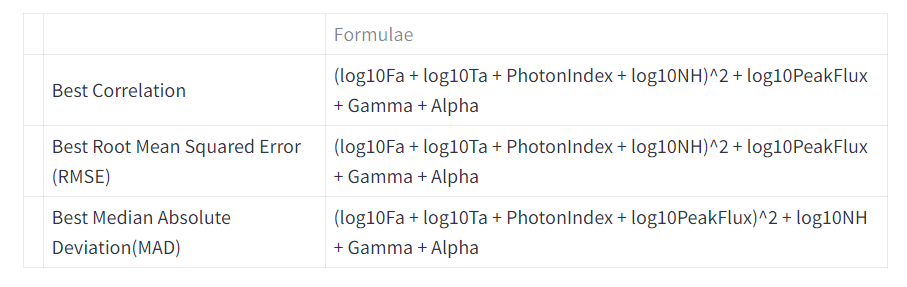}}
    \caption{
        \textit{Image of the GAM and GLM Formula Generator page of the web app. This module allows users to access the best formula for predicting redshift based on the training set provided and its predictors.
    The user can also manually set the cutoff for the correlation and RMSE in order to select the formula.}
    }
    
    \label{fig: App_GetNewFormulas}
\end{figure}

The web app, however, allows for the customization of correlation and RMSE thresholds to determine the formulas that meet the user's criteria.
However, as this work demonstrated, it is advisable to select a higher percentile for correlation and a lower percentile for RMSE.
The idea behind the chosen percentile is that we would need to keep the number of selected formulas small to prevent over-fitting.
The web app determines GAM formulas by default. However, the user can choose GLM by selecting the 'Use GLM' option.
The interface for the web app is presented in the upper panel of Fig. \ref{fig: App_GetNewFormulas} and the final formulas are presented in a tabular format, as presented in the bottom panel of Fig. \ref{fig: App_GetNewFormulas}.

Our approach of using analytic expressions is similar to the symbolic regression approach of \cite{Cranmer2023arXiv230501582C}.
However, the key difference here is that we are not using exponential powers or trigonometric functions of our features.
Given our smaller data sample, we focus on additive expressions. 
However, investigating a symbolic regression could provide an interesting avenue of research. For now, it is beyond the scope of this paper.

\subsection{Outlier removal}\label{outlierremoval}

Before performing any ML training, we need to remove outliers, as these can affect the model performance.
We use the M-estimator technique to remove outliers \citep{huber1996robust,huber1964}.
This is a robust regression method where the outliers are determined based on a robust loss function.
Unlike ordinary least squares regression, which is sensitive to outliers, M-estimator assigns lower weights to extreme observations, making the results more robust.
For the outlier removal, we selected the formula (Eq. \ref{equ}), which obtained the highest counts for GLM.
Using this formula, the M-estimator fits the 238 GRBs and assigns weights ranging from 0 to 1.
GRBs that obtained weights less than 0.65 were eliminated. This specific weight of 0.65 was selected because it removes 5\% of the data set as outliers.
Thus, we were left with 226 GRBs, which were split again into training and test sets containing 180 and 46 GRBs, respectively. 
This training set of 180 GRBs is 20\% larger than the training set used in \cite{Dainotti2024GRBRedshift}.

For the generalization set, we remove GRBs that lie outside the training set's parameter space. 
This minimizes the ML model's chance of extrapolating results, which may lead to inaccurate redshift predictions.
Thus, using our trained model, we predict the redshifts of this trimmed generalization sample of 276 GRBs.

\subsection{The SuperLearner}

SuperLearner is an ensemble method that combines multiple ML models and leverages these models to obtain the best prediction. 
SuperLearner assigns to each model a weight based on its performance (larger weight for better performance), and the sum of all weights is one, {as indicated in the following formula:}
$$
\sum{W_i}=1.
$$
Here, $W_i$ is the weight for the $i^{th}$ model.
The performance of the constituent models is determined via the 10fCV method, which is built into SuperLearner. The advantage of the SuperLearner is that its prediction is a weighted average of the predictions of its constituent models, with better-performing models obtaining larger weights. 
However, different ML models have their own individual implicit assumptions and priors, which can impact the redshift estimation. 
Here, we use the SuperLearner weights to select models whose implicit priors best work with our data set. Models that obtain higher weights are included in the final nested 100 times 10fCV procedure, which provides the results presented in this paper. Furthermore, we are combining parametric (GLM), semi-parametric (GAM) and non-parametric (RF) into the final SuperLearner model, highlighting the benefits of an ensemble approach. A linear model such as GLM enables higher performance for linear relations in the data, while GAM and RF enable a better fit for non-linear relations. Thus, the models share features along with their individual features, enabling a performative ensemble ML model.

\begin{figure}
    \centering
    \fbox{\includegraphics[width=0.95\linewidth]{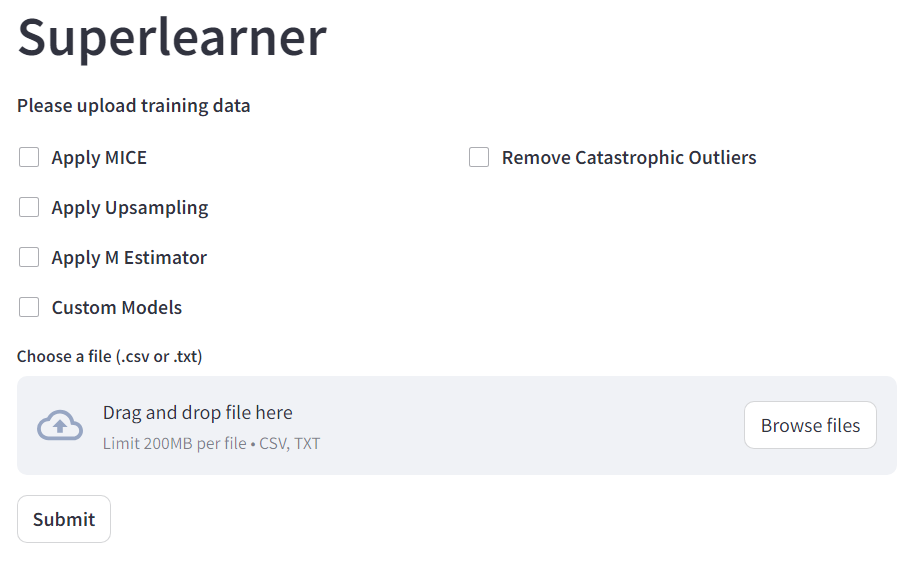}}
    \caption{\textit{Image of the SuperLearner page of the web app. Shown are the data upload panel and options to choose model and data parameters.}
    }
    \label{fig: App_SuperLearner1}
\end{figure}

In our web app, we allow users to provide their data and obtain results of the SuperLearner through the metrics of our analysis: the Pearson correlation plots, both in linear and log scale, between $z_{obs}$ and $z_{pred}$, the bias, defined as $<z_{pred}-z_{obs}>$, the normalized median absolute deviation (NMAD), and the RMSE. 
Additionally, the SuperLearner module of the web app also supports features such as M-estimator, MICE, the removal of catastrophic outliers, model integration such as GAM, GLM, etc., and parallel processing.
This enables users to tune the functionality to their liking.
The user interface of the Superlearner module in the web app is presented in Fig. \ref{fig: App_SuperLearner1}.

The results presented here use a SuperLearner model that incorporates six GAM formulas, two smooth GAM formulas, four GLM formulas, and RF (see Sec. \ref{testedmodels} and  Sec. \ref{formulagen} for details).

\subsection{Balance sampling}\label{balancedsampling}
We aim to investigate how a more balanced sample of sources will affect the performance of our ML model. 
A balanced sample is a data set in which the response variable's distribution is close to uniform. 
Such a distribution helps train the model over the full range of potential values for the response variable, thus helping it generalize to future samples.
In this analysis, we deal with an unbalanced sample where the number of GRBs at $z>2.5$ (80) is fewer than the number of GRBs at $z<2.5$ (158). 
This phenomenon can be attributed to the Malmquist bias effect, which causes us to only detect the brightest astrophysical sources at high-$z$ due to detector thresholds.
This effect prevents us from seeing a large population of dim sources at high-$z$. 
If we had a perfect satellite with extremely high sensitivity, we could populate the distribution at high-$z$. 
There are several methods to balance the sample: up-sampling, down-sampling, and balance sampling.
For up-sampling, the original data remains unchanged, and duplicate samples of the minority class are added.
In down-sampling, data are randomly selected to ensure that all classes are equally represented, matching the frequency of the minority class.

For up-sampling, multiple methods exist. We applied the Synthetic Minority Over-sampling Technique (SMOTE, \cite{SMOTE,ubBalance}).
As the name implies, SMOTE creates synthetic copies of the minority class, thereby increasing its sample count. This is achieved by generating new data points within the feature space of the minority class dictated by the real nearest-neighbor data points.
This prevents overfitting, which can happen when minority class samples are duplicated. 
It also eliminates the need to remove samples from the majority class, which may lead to information loss and reduced accuracy.

\begin{figure}
    \centering
     \includegraphics[scale=0.49]{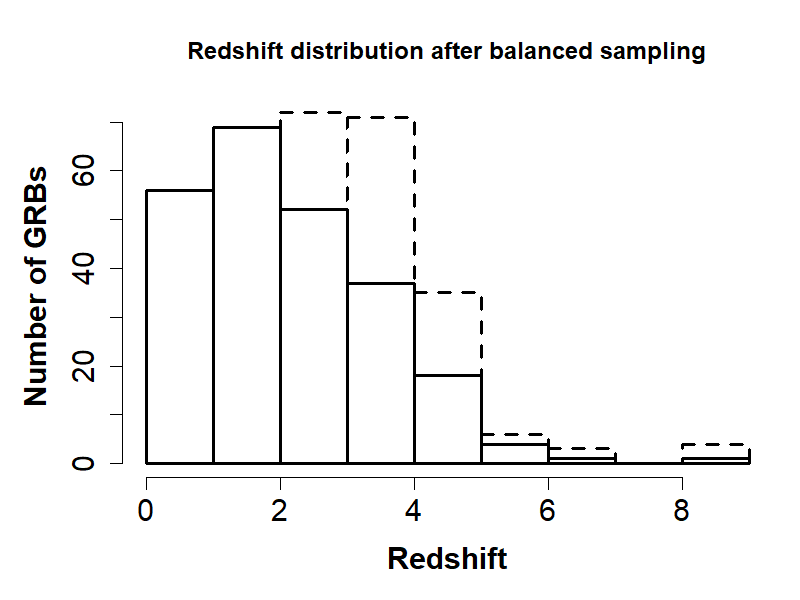}
    \caption{\textit{Redshift distribution obtained with the upsampling (solid lines) and the original redshift distribution (dashed lines).}
    }
    \label{fig:redshift distribution}
\end{figure}

The redshift distribution of the upsampled data are shown in Fig. \ref{fig:redshift distribution}.
This balanced sample also undergoes the same data treatment process as mentioned in Sec. \ref{outlierremoval}.

Since this is an important part of the inference of the redshifts, we have also made this part available through the web app, see Fig. \ref{fig: App_SuperLearner1}. 
The user can tick the 'Apply Upsampling' button to obtain results with an augmented and balanced sample.

\subsection{Bias correction}\label{BC}

Bias correction is a technique that is used to correct biases in an ML model's prediction.

Bias is defined as the difference between the observed and predicted response variables.
ML models trained on an imbalanced training set are more biased toward the range containing more examples.
In our sample of GRBs, the models tend to predict more toward lower redshift ranges as there are more GRBs in that range (see Sec. \ref{balancedsampling}).
Balanced sampling can mitigate this issue to an extent. However, we applied the optimal transport bias correction technique \citep{optimaltransport} to address this issue more directly.
Some of us have already used it in \cite{Dainotti2024GRBRedshift,Dainotti2024ApJ...967L..30D,Narendra2022}.

In this technique, first, the observed and predicted values of the response variable are sorted in ascending order.
Next, a linear model is fit between them and the predicted values are corrected by multiplying the slope ($A)$ and adding the intercept ($B$) of the linear fit.
Furthermore, bias correction is performed separately in three redshift ranges: $0<z<2$, $2<z<3.5$, and $3.5<z<8.4$. 
These particular ranges were selected as they produced the best results for our particular ML model and data set.
The bias correction equation is as follows:
$$
Z_C=B + A*Z_P,
$$
where $Z_C$ is the corrected $\log(z+1)$ prediction, $Z_P$ is the SuperLearner prediction.
The assumption we are making here is that the $\log(z_{pred}+1)$ follows the same distribution as $\log(z_{spec}+1)$.

\subsection{Monte Carlo analysis}\label{MCMC}

{With the MC simulations, we aim to explore the reason for the changes in our results compared to \cite{Dainotti2024GRBRedshift}.
{The MC simulations also help us assess the degree of uncertainties in the prediction once the sample size is increased. 
In the procedure detailed in \cite{Dainotti2024GRBRedshift}, we applied MICE to impute the missing data and then ran Superlearner on the imputed data. 
Therefore, in this work also we perform MC simulations on the imputed data to keep the comparison more straightforward.}
}

{We perform simulations on the 7 predictors which are selected by LASSO, {see Fig. \ref{fig:lasso}. }
{We simulate three sample sizes: 14, 28, and 42. This reflects, on average, how many GRBs with redshift and plateau will be added to the Swift catalog in one, two, and three years, respectively. 
This average was calculated based on the training set currently being used.
Each sample size was generated 200 times with unique instances and combined with the training set of observed GRBs. 
For sampling, we used the SMOTE method (see Sec. \ref{balancedsampling}), as this enables multivariate sampling of the data. 
Multivariate sampling is important as it maintains the correlations between the seven parameters involved, such as between $\log(T_a)$ and $\log(F_a)$ and $\log(Peakflux)$ \citep{Dainotti2008}.}
We simulated the samples so that the distributions of the variables resemble the current ones, and thus we retained only the samples that pass the KS test.}
{Each simulated sample was tested with 10fCV using the ensemble SuperLearner model. The results from this analysis are presented in Sec. \ref{mcmcresults}.}

\subsection{Model comparison with web app}

\begin{figure}[h!]
    \centering
    \fbox{\includegraphics[width=0.49\textwidth]{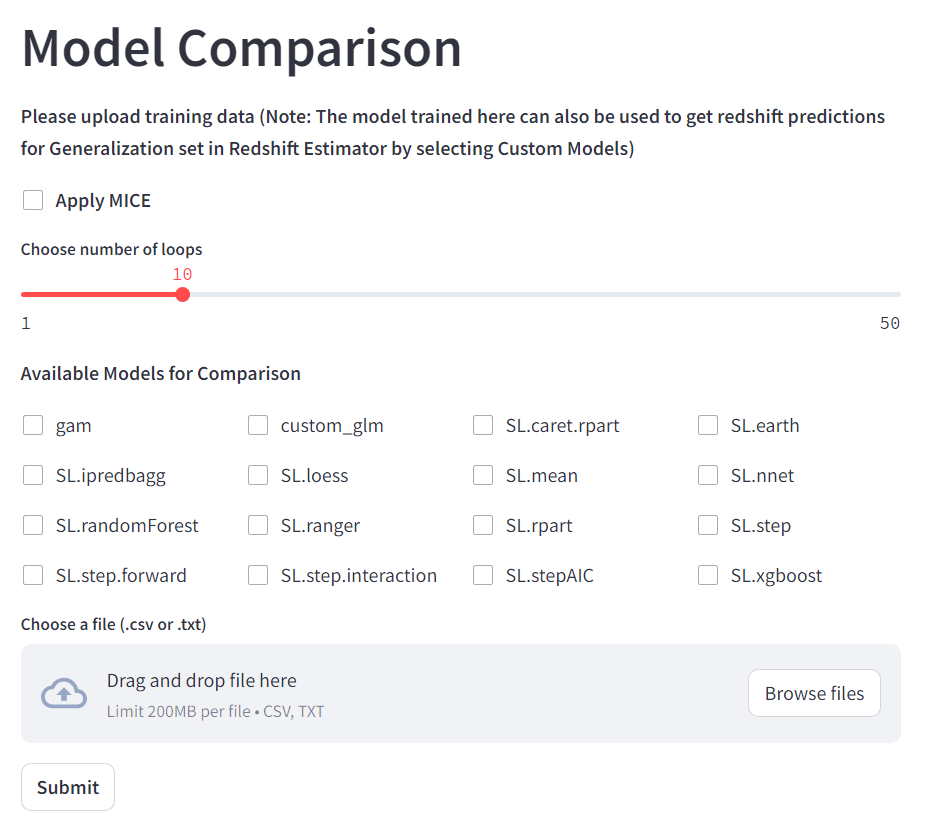}}
    \caption{\textit{Various options and parameters available to a user of the web app when using the Model Comparison functionality.}}
    \label{fig: App_Model}
\end{figure}

The web app's Model comparison module is designed to allow users to investigate their data using more ML models than those used in this work.
This is important because the best model for predicting redshift may vary depending on the data being used.
Figure \ref{fig: App_Model} shows the Model Comparison module's interface on the web app.
This process is initiated by importing data files provided by the user. 
Moreover, this module also offers functionalities such as MICE for data imputation and allows users to specify the number of loops for 10fCV (see Sec. \ref{formulagen}).
The web app lists predefined models for use, as shown in Fig. \ref{fig: App_Model}. 
Users can tailor their selections based on specific requirements and commence the analysis by clicking on the submit button. 
The application then constructs an ensemble SuperLearner model based on the selected parameters and models.
The application generates a correlation plot depicting the $z_{pred}$ vs $z_{spec}$ distribution and metrics showing the performance of the selected models over the provided data set. 
This visualization aids users in understanding the relative strengths and weaknesses of the models under consideration.

\subsection{Relative importance of variables}

The Relative Importance module of the web app evaluates the relative significance of the variables used for predicting the redshift.
The process is initiated by importing data files the user provides. 
The ML model used for this module is the best model determined from our analysis in this publication (see Sec. \ref{testedmodels}). 
The relative importance of the predictors is decided based on this model's performance.
Users can also use the pretrained models from the web app's Model Comparison or Superlearner modules to obtain relative importance values for their data.

\begin{figure}[h!]
    \centering
    \fbox{\includegraphics[width=\linewidth]{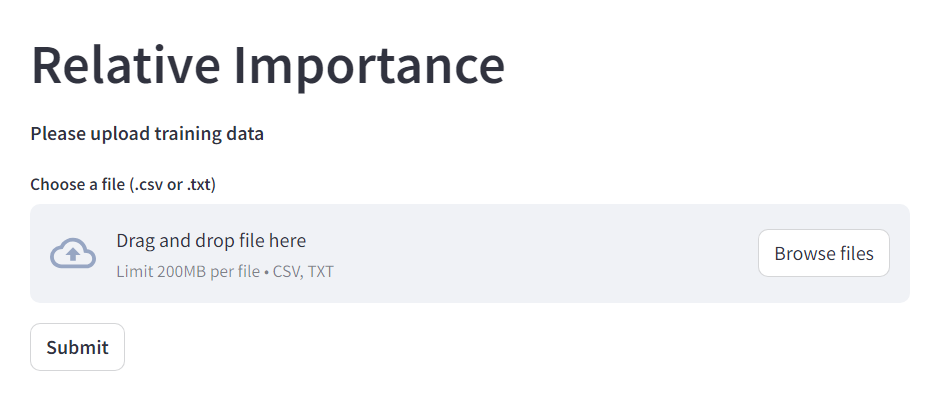}}
    \fbox{\includegraphics[width=\linewidth]{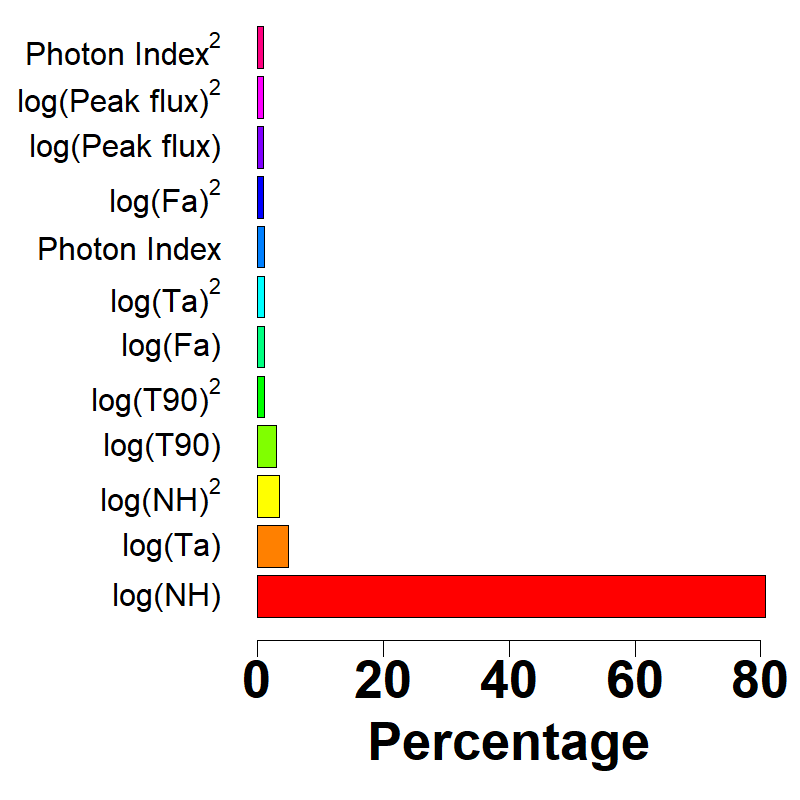}}
    \caption{\textit{
    Upper panel: User interface of the Relative Influence module of the web app. Lower panel: Relative importance bar plot for the predictors used in this analysis.}}
    \label{fig: App_RelImp}
\end{figure}

The interface of the web app for this module is presented in the upper panel of Fig. \ref{fig: App_RelImp}.

\subsection{Data visualization with the web app}

The Data Visualization module of the web app provides the user with an option to visualize their training set or generalization set via the scatter matrix plot (as shown in Fig. \ref{fig:scattermatrix}) and missing data (as shown in Fig. \ref{fig:mice}).
The scatter matrix plot offers a comprehensive view of the uploaded data along with histograms and 
correlation coefficients between pairs of features. 
This visualization tool facilitates a deeper understanding of the data set's relationships and distributions, aiding users in making informed decisions and gaining valuable insights.
This module's plot of missing data helps users determine which features in the provided data set have missing entries.

\begin{figure}[h!]
    \centering
    \fbox{\includegraphics[width=0.95\linewidth]{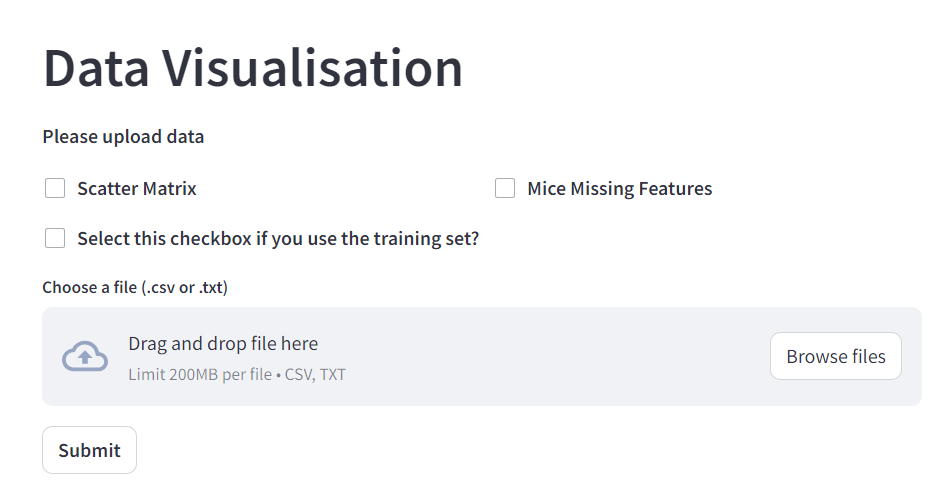}}
    \caption{\textit{Interface for the data visualization module of the web app.}}
    \label{fig:datavisualization}
\end{figure}

\section{Results}\label{results}

Our Results section is split into four subsections.
First, we present the results demonstrating the effectiveness of our ML model. Here, we also show the results obtained using balanced sampling.
Second, using the trained model, we present the redshift predictions obtained on the generalization set.
Third, we present the results from our MC analysis.
Finally, we provide an overview of the performance of the web app's redshift estimator module.

\subsection{Performance of the machine learning model}\label{mlresults}

\begin{figure*}
    \centering
    \includegraphics[width=0.49\linewidth]{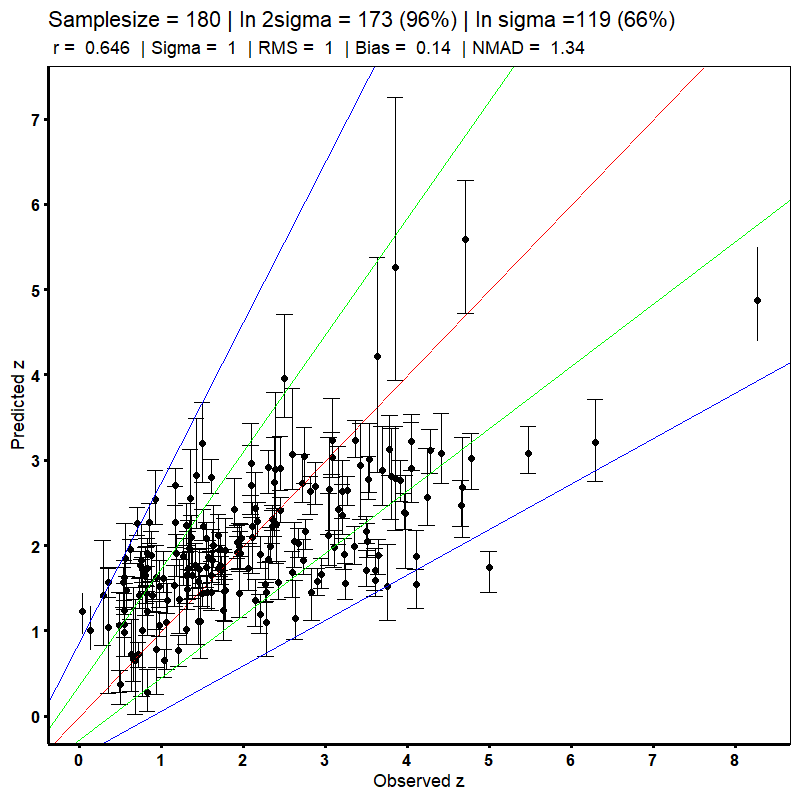}
     \includegraphics[width=0.49\linewidth]{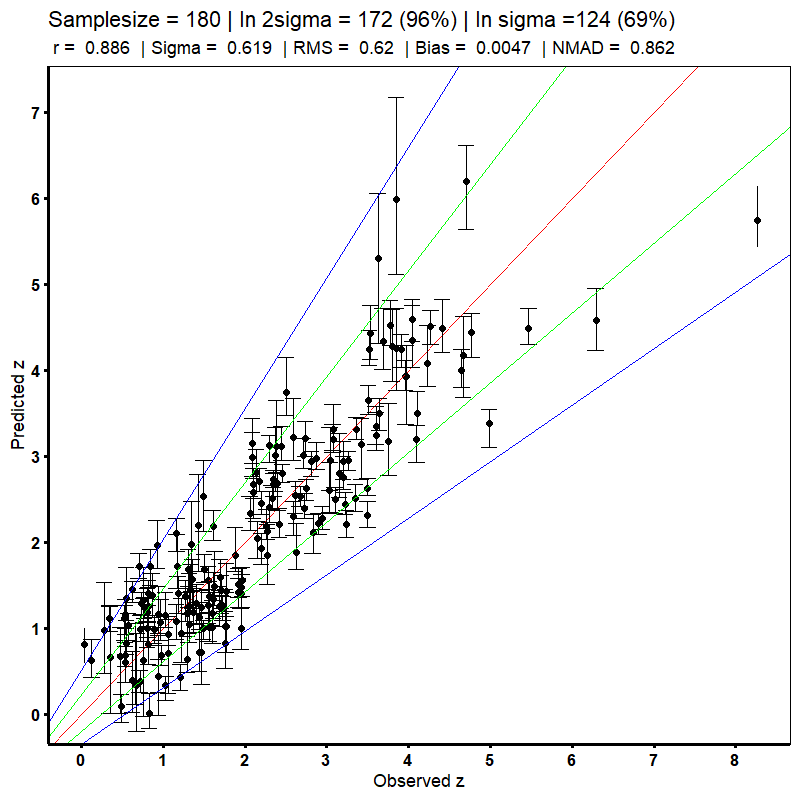}
     \includegraphics[width=0.49\linewidth]{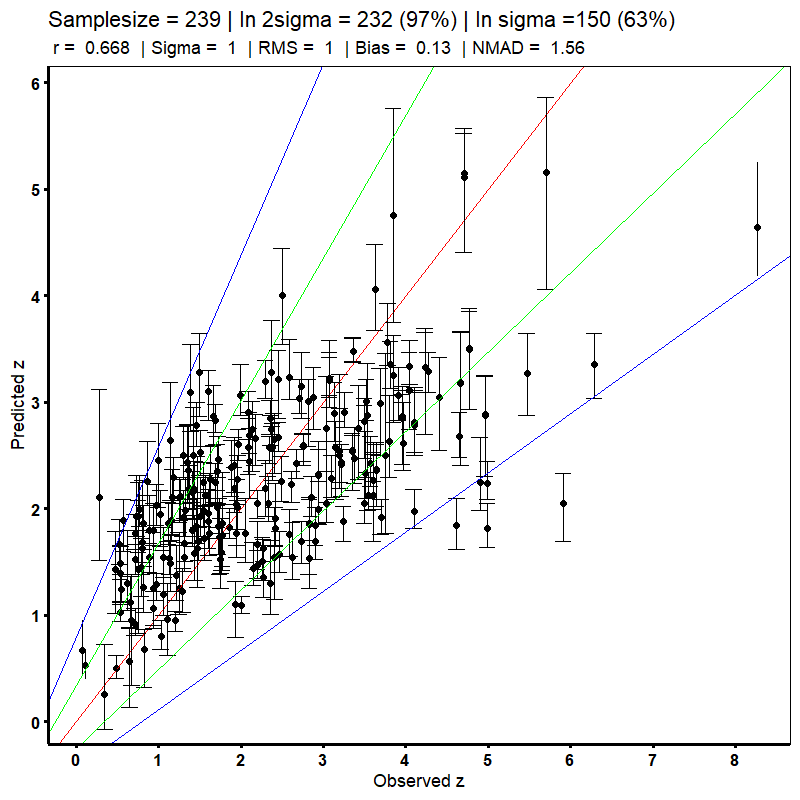}
     \includegraphics[width=0.49\linewidth]{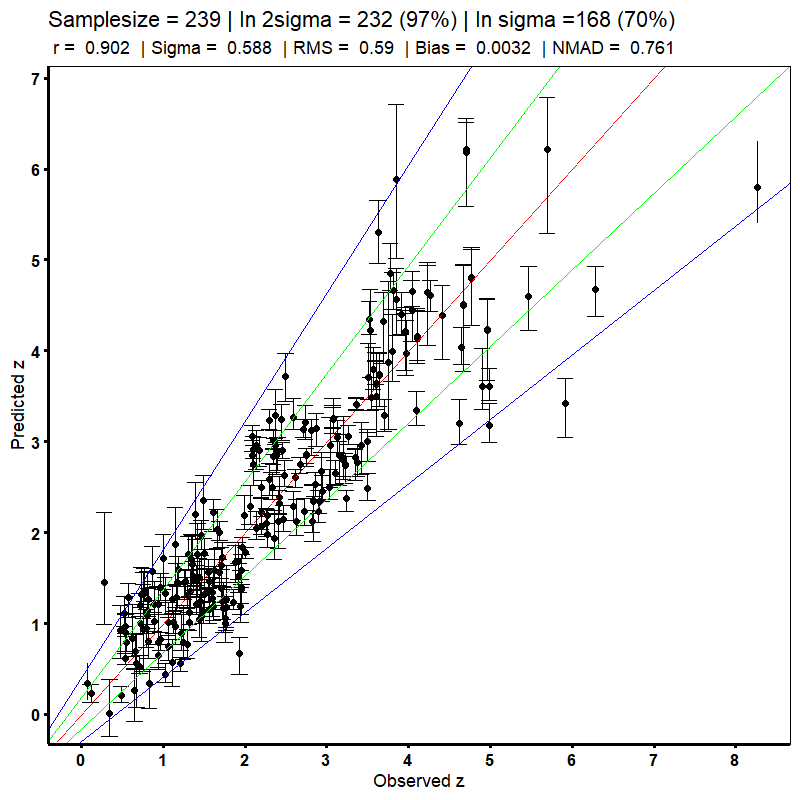}
    \caption{\textit{Results obtained from performing 10fCV 100 times using the SuperLearner model. The bottom and upper panels present results with and without balanced sampling, respectively.
    The upper-left panel shows the results obtained from the SuperLearner model. The upper-right panel shows the corresponding bias-corrected results.
    The bottom-left panel shows the result obtained from the SuperLearner model when using a balanced sample. The bottom-right panel shows the corresponding bias-corrected results.}
    }
    \label{fig:superlearner_results}
\end{figure*}

The results shown here are obtained after performing 10fCV 100 times on 180 GRBs using the SuperLearner model.
The correlation plot between $z_{obs}$ and $z_{pred}$ obtained from SuperLearner is shown in the upper left panel of Fig. \ref{fig:superlearner_results}. 
We achieve an $r$=0.646, RMSE=1.011, bias=0.14 and NMAD = 1.34.
We apply the optimal transport bias correction (see Sec. \ref{BC}) to these results to obtain the plot shown in the upper right panel of Fig. \ref{fig:superlearner_results}.
Here, we obtain $r$=0.89, RMSE = 0.62, bias = 0.0047, and NMAD = 0.86.
Sec. \ref{discussion} discusses how these results compare with those presented in \cite{Dainotti2024GRBRedshift} and its implications.

We also present the results for our ML model when using a balanced sample.
The plots are presented in the bottom left and right panels of Fig. \ref{fig:superlearner_results}.
The bottom left panel shows the correlation plot obtained from the SuperLearner model, similar to the upper left panel without balanced sampling.
We achieve $r$ = 0.668, RMSE = 1.005, bias = 0.13, and NMAD = 1.56.
The results then undergo the same bias correction step, and they are shown in the bottom right panel of Fig.\ref{fig:superlearner_results}. 
Here we obtain an $r$= 0.902, RMSE= 0.59, bias= 0.003, and NMAD= 0.76.
The discussion about these results compared to those without balanced sampling is presented in Sec. \ref{discussion}.

{\cite{ukwatta2016machine} obtained $r=$0.57 using RF. We have the four features common with their analysis, namely $T_{90}$, $PhotonIndex$, $\log(Fluence)$, and $\log(NH)$. 
Our correlation value is} 13.3\% higher for the SuperLearner results and 56\% higher for the bias-corrected results.
Our balanced sample results are 17.1\% higher, and the bias-corrected results are 58\% higher.
We can attribute this improvement to our use of a more robust methodology and the plateau features. 
We also obtained similar results in \cite{Dainotti2024GRBRedshift}.

\cite{Aldowma2024MNRAS.529.2676A} use DeepNeural networks and RF with Fermi-GBM and Konus-Wind observed GRBs for estimating pseudo-redshifts.
They achieved the highest $r^2$ of 0.86 on their test set, which is equal to $r=$0.92.
Compared to our bias-corrected results, they achieved an improvement of 4\%.
However, in their analysis, it is unclear whether they discard duplicated GRBs between the Konus-Wind and Fermi-GBM catalogs. Furthermore, they seem to have trained the model on a single training-test split. These two effects can make the ML model overfit their data, leading to a higher correlation score. 

{\cite{Racz2023CoSka..53d.100R} used RF and extreme gradient boosting with GRBs taken directly from the Swift catalog. 
Similar to our analysis, they use $\log(z+1)$ as their response variable and obtain a $r=$0.76 in the $\log(z+1)$. However, other metrics such as RMSE, bias, and NMAD are not quoted. Furthermore, they do not quote the metrics in the linear scale.
Thus, comparing our results in the $\log(z+1)$ scale, we find that our $r$ is lower by} 20\% and higher by 19\% for the non-bias-corrected and bias-corrected results, respectively.

\subsection{Relative importance results}\label{relimp_result}

{The relative importance of the seven features is presented in the lower panel of Fig. \ref{fig: App_RelImp}.
Similar to \cite{Dainotti2024GRBRedshift,Dainotti2024ApJ...967L..30D}, we observe that the highest influence on the redshift prediction is assigned to $\log(NH)$.
The $\log(NH)$ is obtained from the Swift catalog and is determined by spectral fitting with a fixed galactic absorption with the HEASoft software\footnote{https://heasarc.gsfc.nasa.gov/ftools}. 
Thus, the $\log(NH)$ value we use here is the column density between the burst and the Milky Way.} 

The dependence of $\log(NH)$ on the redshift is expected; however, this relation is non-trivial. The value of $\log(NH)$ is impacted by the density and metallicity of the intervening medium, and thus, we discuss some of the factors that can affect this complex correlation\citep{Arumaningtyas2024}.

\begin{itemize}
    \item The first factor is the absorption of the GRB's emission by the intergalactic medium between our galaxy and the GRB's host galaxy. 
    Evidence for this was found by \cite{Rahin2019}, who attributed the evolution of $\log(NH)$ with redshift to the accumulation of the ionized and diffuse gas in the intergalactic medium.
    \item The second factor is the interstellar medium (ISM) density inside the host galaxy. 
    Generally, high-$z$ galaxies have larger star-forming regions, and this corresponds to the presence of more neutral hydrogen in the ISM.
    This causes more absorption of the GRB's emission than inside low-$z$ galaxies, thus affecting the $\log(NH)-z$ relation.
    \item The third factor is the evolution of the metallicity of the progenitor's ejecta. \cite{Heintz2023} found that the effective metallicity derived from the GRB spectra does evolve with redshift, and thus, it might have some impact on the $\log(NH)-z$ relation.
    We expect the progenitors of GRBs at high-$z$ to have lower metallicity than their low-$z$ counterparts.
    This can lead to the complex evolution of absorption of the GRB's emission and thus significantly affect the $\log(NH)-z$ relation.
    \item Finally, the metallicity of the host galaxy's ISM has a limited impact on the $\log(NH)-z$ relation. 
    This is evident from the findings of \citep{Cucchiara2015,Graham2023ApJ...954...13G} where they observed that high-$z$ GRBs occur in low metallicity regions. 
    Furthermore, \cite{Bolmer2018A&A...609A..62B} found that GRB host galaxies at $z>4$ have less dust and extinction value ($Av<0.5$mag) than low-$z$ sources. This leads to the conclusion that the evolution of metallicity cannot be explained by higher dust content at $z>4$.
\end{itemize}

Going back to the relative importance, then the second most predictive feature here, and in \cite{Dainotti2024GRBRedshift}, is $\log(T_a)$. The plateau features are derived using the X-ray LCs in both cases.
This differs from the second most predictive variable found in \cite{Dainotti2024ApJ...967L..30D}, which was $\log(F_a)$. The plateau features were obtained from optical LCs there.
The fact that $\log(T_a)$ is correlated to $\log(F_a)$ implies that the difference between the X-rays and optical results is indeed not significant.
The discussion about the implication of these results is presented in Sec. \ref{discussion}.

\subsection{Estimating new redshifts in the generalization sample}\label{generalization_results}

\begin{figure*}
    \centering
    \includegraphics[width=\linewidth]{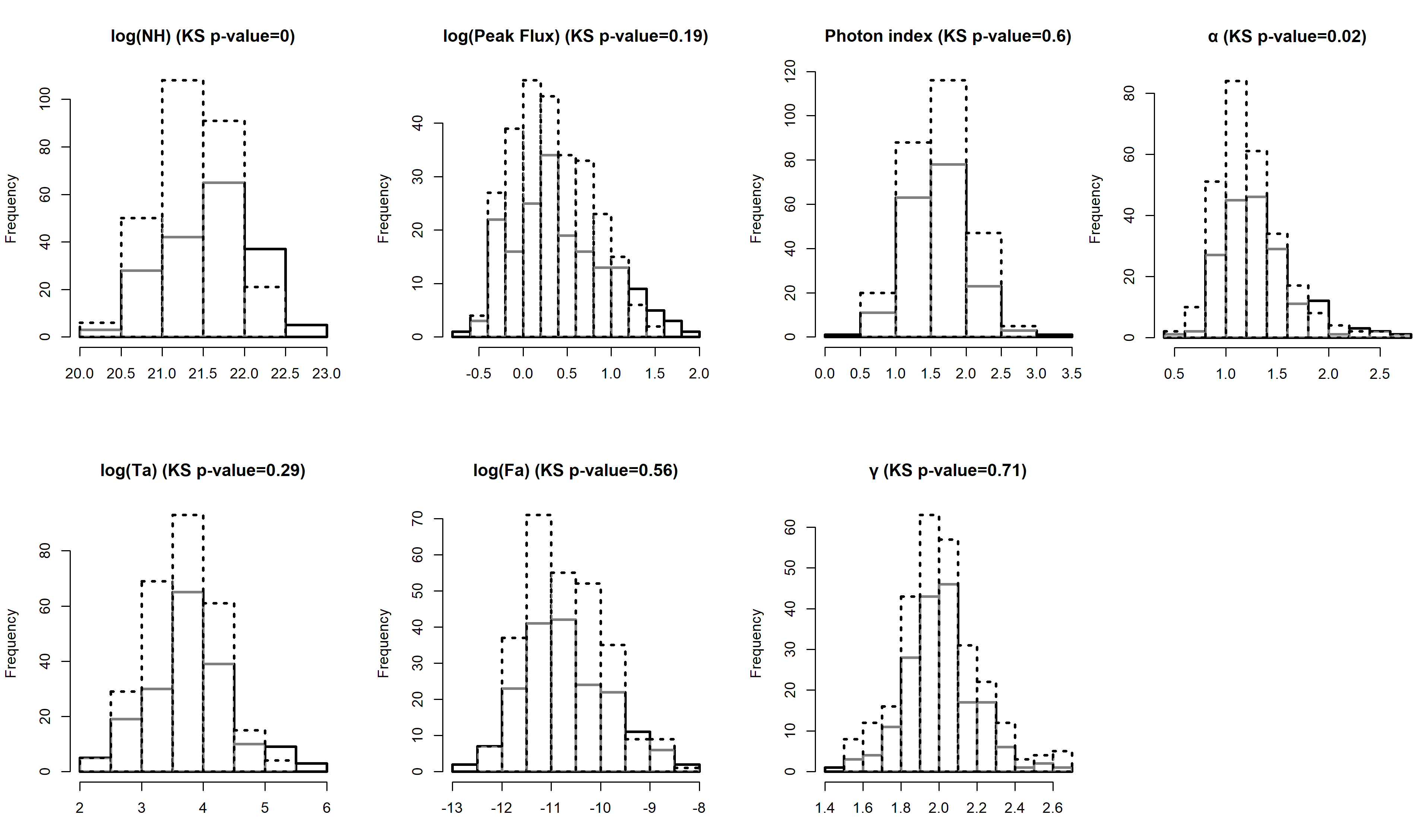}
    \caption{\textit{Histograms showing the distribution of the seven features of the generalization set against the training set. The KS test p-value scores are shown in the title of each figure. The dashed lines and solid lines represent the generalization and training sets, respectively.}}
    \label{fig:genset_distribution}
\end{figure*}

\begin{figure*}
    \centering
    \includegraphics[width=\textwidth]{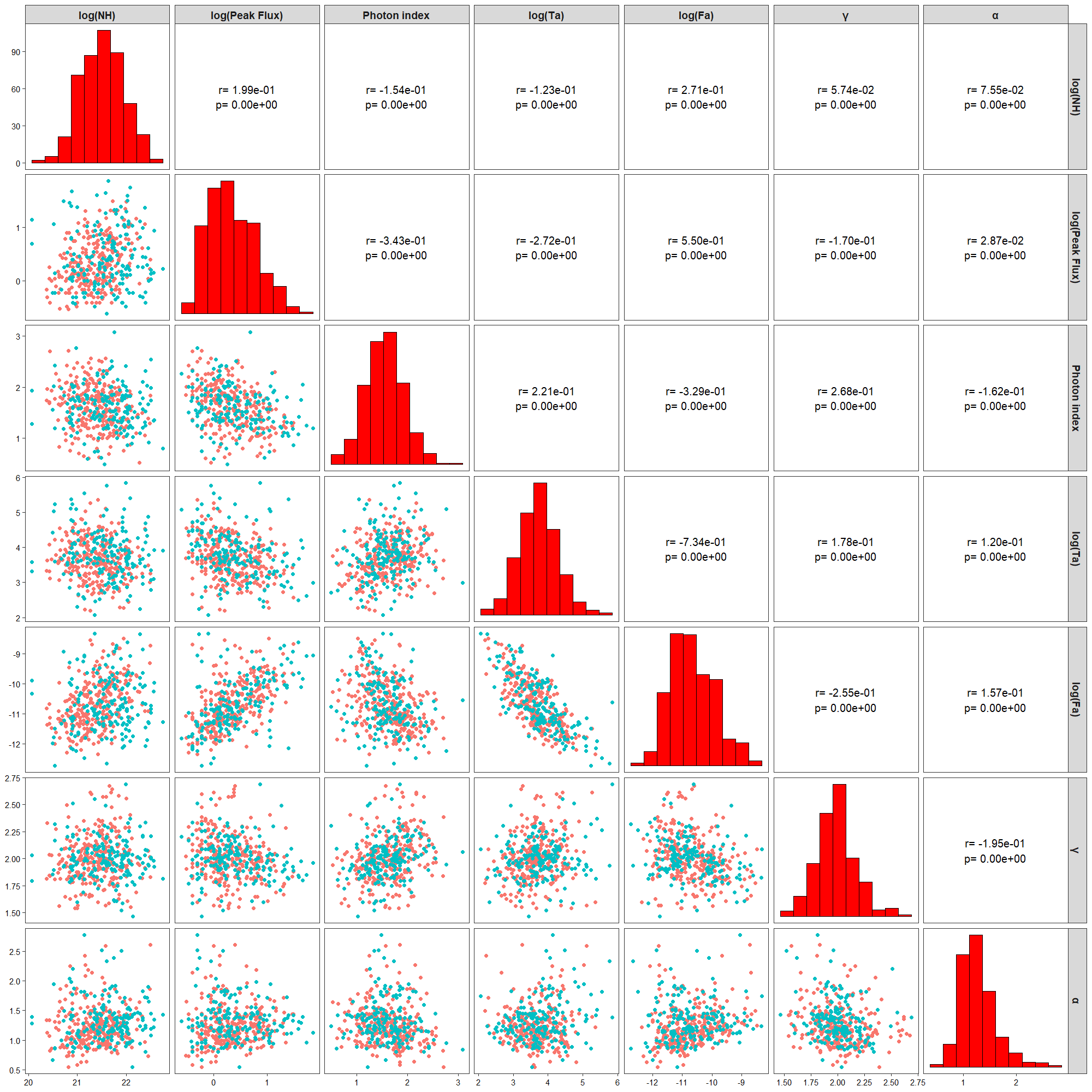}
    \caption{\textit{Scatter matrix plot of the generalization and training sets. 
    The generalization set is presented with orange dots, and the training set is presented with cyan dots.
    The diagonal shows the histogram distribution for the generalization set, while the upper triangle shows the Pearson correlation values for respective pairs of parameters.}
    }
    \label{fig:gen+training_scatter}
\end{figure*}

As stated in Sec. \ref{datasample}, we have increased the generalization sample by 35\%, from 221 GRBs \citep{Dainotti2024GRBRedshift} to 276 GRBs.
The histogram and scatter matrix plots comparing the training and generalization sets' features are shown in Fig. \ref{fig:genset_distribution} and Fig. \ref{fig:gen+training_scatter}, respectively.
We compared the distribution of each parameter between the generalization set and the training set via the KS test, and the p-values are stated at the top of each parameter distribution.

Compared to the training set, all distributions, except the one of $\log(NH)$, have a p-value greater than 0.05. 
This means that the features are sampled from the same parent distribution. 
This indicates that the training set's distribution of $\log(NH)$ differs from that of the generalization set and is not representative of the parent population.

\begin{figure}
    \centering
    \includegraphics[width=0.99\linewidth]{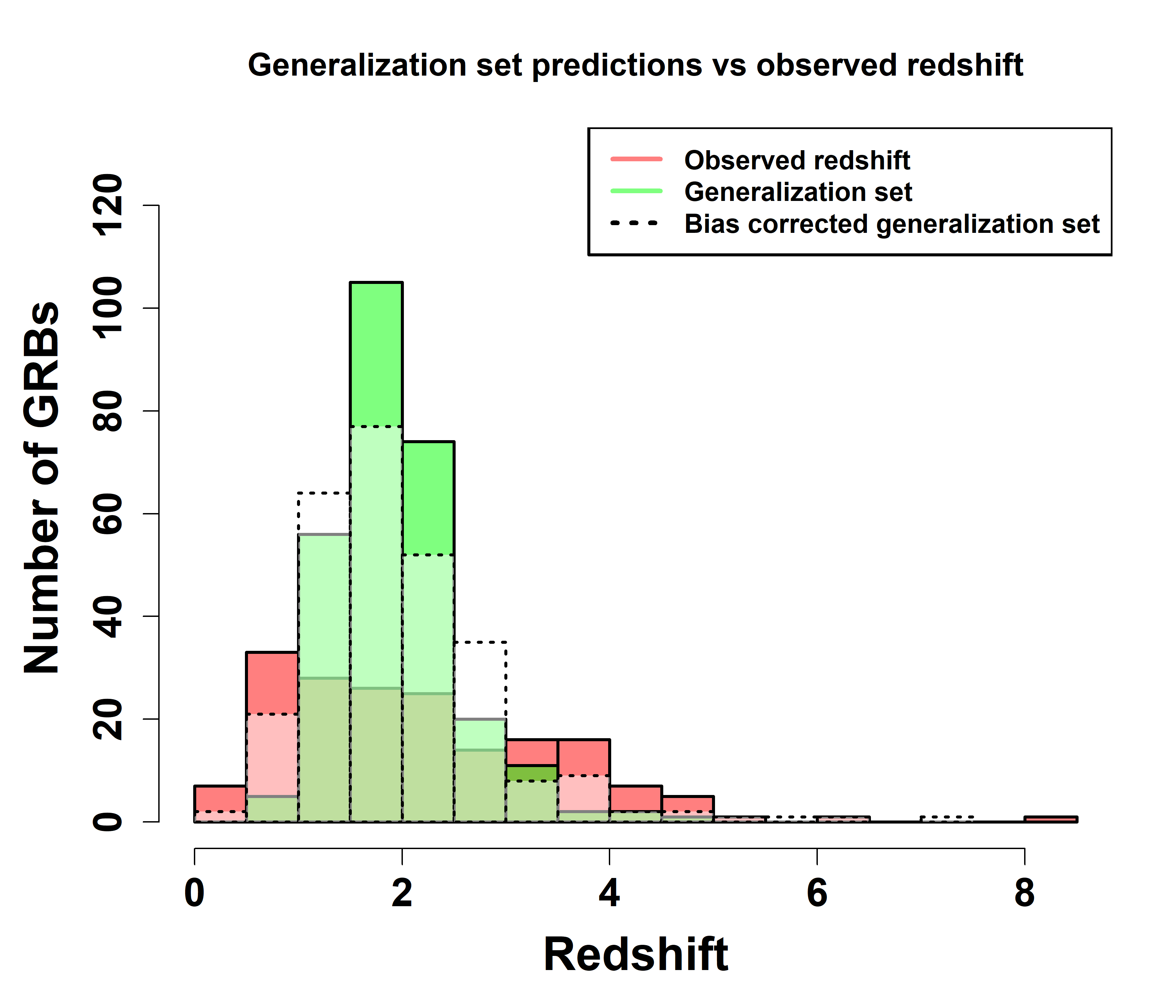}
    \includegraphics[width=0.99\linewidth]{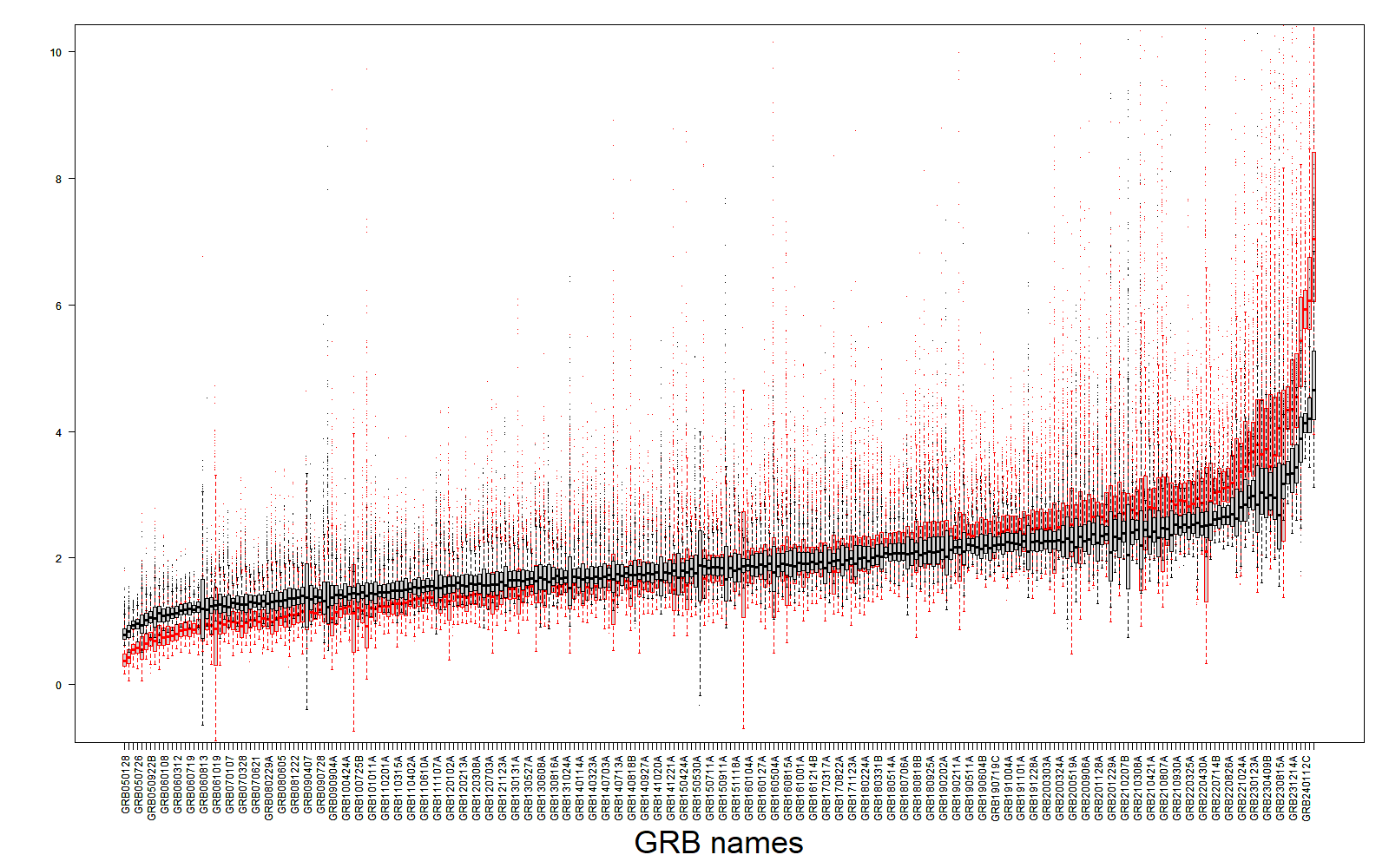}
    \caption{\textit{Upper panel: Redshift distribution of the generalization and training sets. The training set is shown with red histograms, while the generalization set with and without bias correction is shown in white and green histograms, respectively.
    Lower panel: Box plot distribution of the generalization set redshift predictions, sorted from lowest to highest redshift.
    The bias-corrected and non-bias-corrected distributions are shown in red and gray box plots, respectively. The dots show the outlier predictions obtained.}
    }
    \label{fig:generalizationset_results}
\end{figure}

In Fig. \ref{fig:generalizationset_results}, we compare the distributions of predicted redshift and the redshift of the training set.
The training set's redshift distribution is presented in red histograms, and the generalization set's redshift distributions are presented in white and green histograms, with and without bias correction, respectively. 
Performing the KS test between the redshift distributions of the training set and the generalization set gives p-values lower than the threshold of 0.05. 
Specifically, we obtain p-values of $2.5\times10^{-7}$ and $0.0002$ for the non-bias-corrected and bias-corrected redshift distributions, respectively.
This discrepancy can be attributed to the different $\log(NH)$ distributions of the generalization and training sets.
However, we can observe that the bias correction step does bring the redshift distribution closer to the training set distribution. Thus, this is an essential step in the redshift prediction.

It should also be noted that GRBs with and without $z_{obs}$ might differ in their physical origin and chances of observations. 
Thus, they can have different intrinsic redshift distributions.
The low-luminosity sources have a lower chance of having their redshift measured due to the difficulty of obtaining the redshift from the afterglow spectral features, the trouble of observing the host galaxy, or the fact that the host galaxy is not in the database and is not pointed by any observing facilities.
Moreover, we observe a statistically smaller $\log(NH)$ in the generalization set. 
This corresponds to less absorption via the intergalactic medium, thus making the absorption-based redshift measurements harder. 
Those two effects could explain the difference in redshift distributions between the training and generalization sets.
{
We discuss the physical origins of each variable in \cite{Dainotti2025ApJ...978...51D}.
However, we can adopt certain strategies to help us constrain the differences. 
The first strategy one could adopt would be to ensure that the parameter spaces of the generalization set and training set are same, which is what we do here in this paper.
{Indeed, to prevent biases in the pseudo-redshifts, we trim the generalization set before predicting its redshift so that our ML model will predict only those GRBs whose parameter space falls within our training set.} 
This will ensure that any outlier GRBs in the generalization set that are more likely to have a different physical origin can be minimized.
Another strategy would be better classify GRBs, regardless of whether they have $z$ or not.
Indeed, some of us have already attempted to address this \cite{Bhardwaj2023MNRAS.525.5204B}, where we attempted to classify GRBs based on their properties using unsupervised ML.
Efforts in this direction can help ensure that a homogeneity is maintained between the training set of an ML model and the generalization set used.
Next, we have already investigated how the performance of our ML model changes when more GRBs are added to the generalization set, making sure that our model is predictive enough to handle the additional data and continue to perform at the expected level.
We explore this in the next section.
}

We show a box plot distribution of the predicted redshifts in the lower panel of Fig. \ref{generalization_results}.
The red and gray box plots show the bias-corrected and non-bias-corrected distributions.
The x-axis shows the names of the generalization set GRBs, and the y-axis shows the redshift values.
We note that because of the image's scale, not all the GRB names are present along the axis.
{Our work here has enabled us to increase the Swift sample with pseudo-redshifts by 110\%.}

\subsection{Monte Carlo simulation results}\label{mcmcresults}
\begin{figure*}
    \centering
    \includegraphics[width=0.31\linewidth]{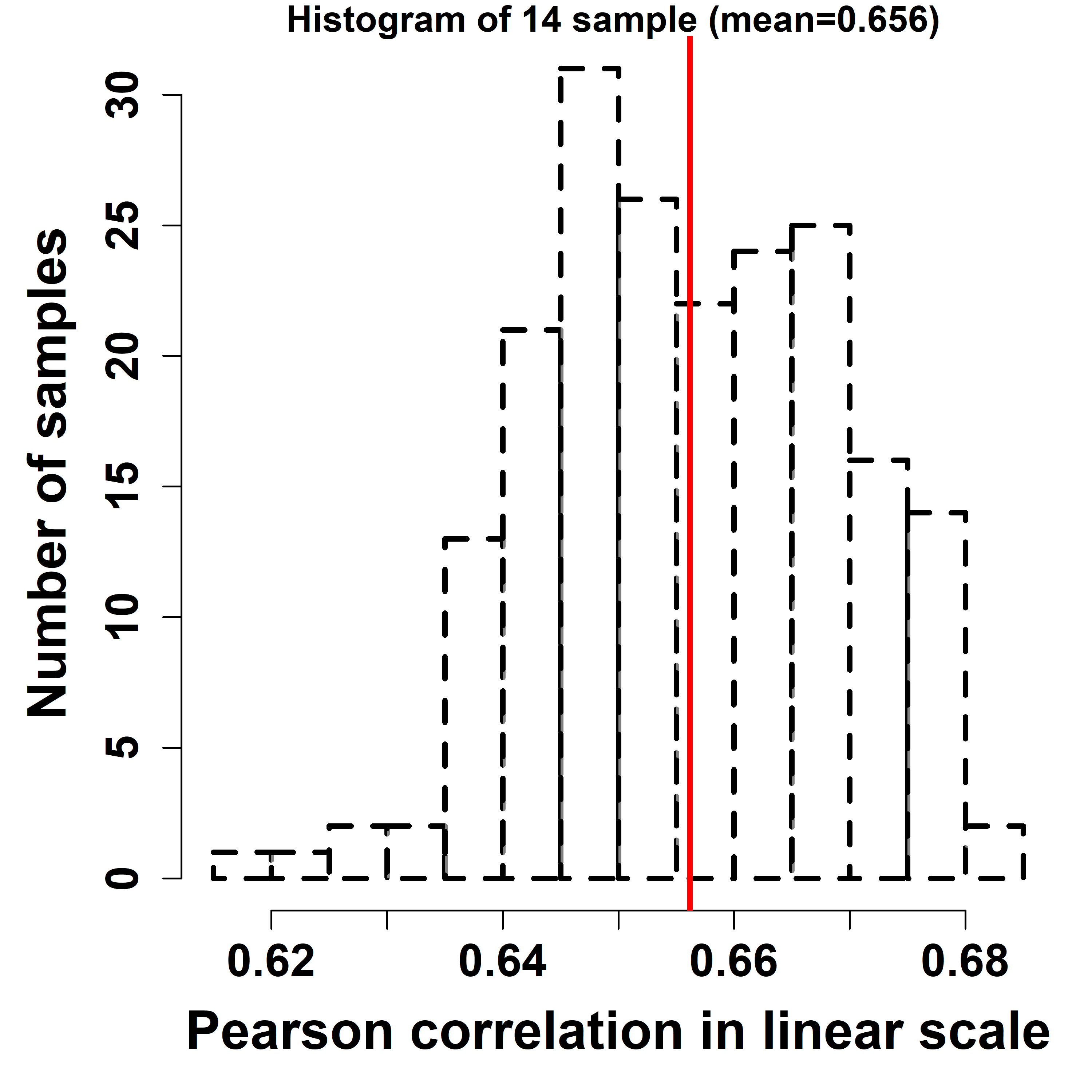}
    \includegraphics[width=0.31\linewidth]{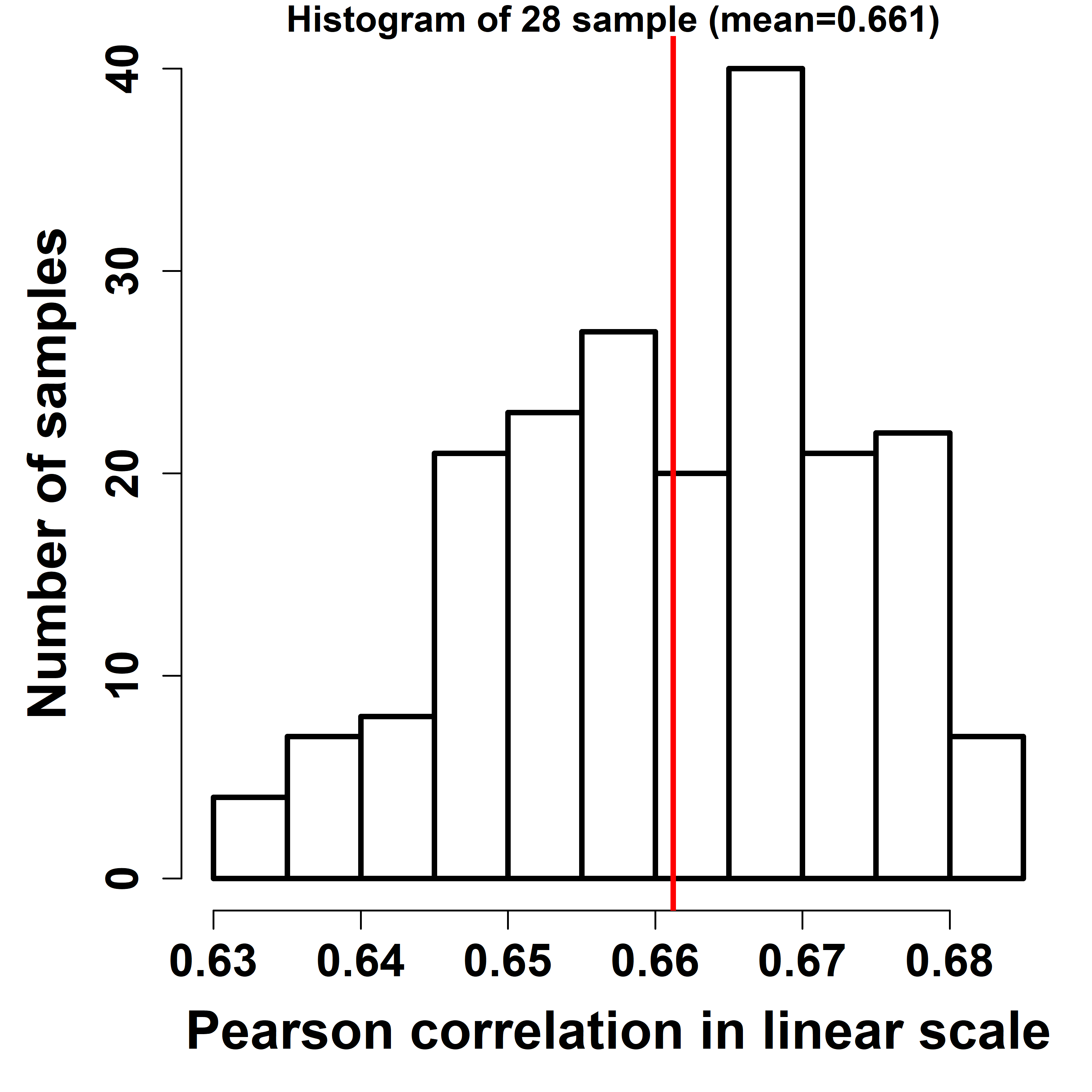}
    \includegraphics[width=0.31\linewidth]{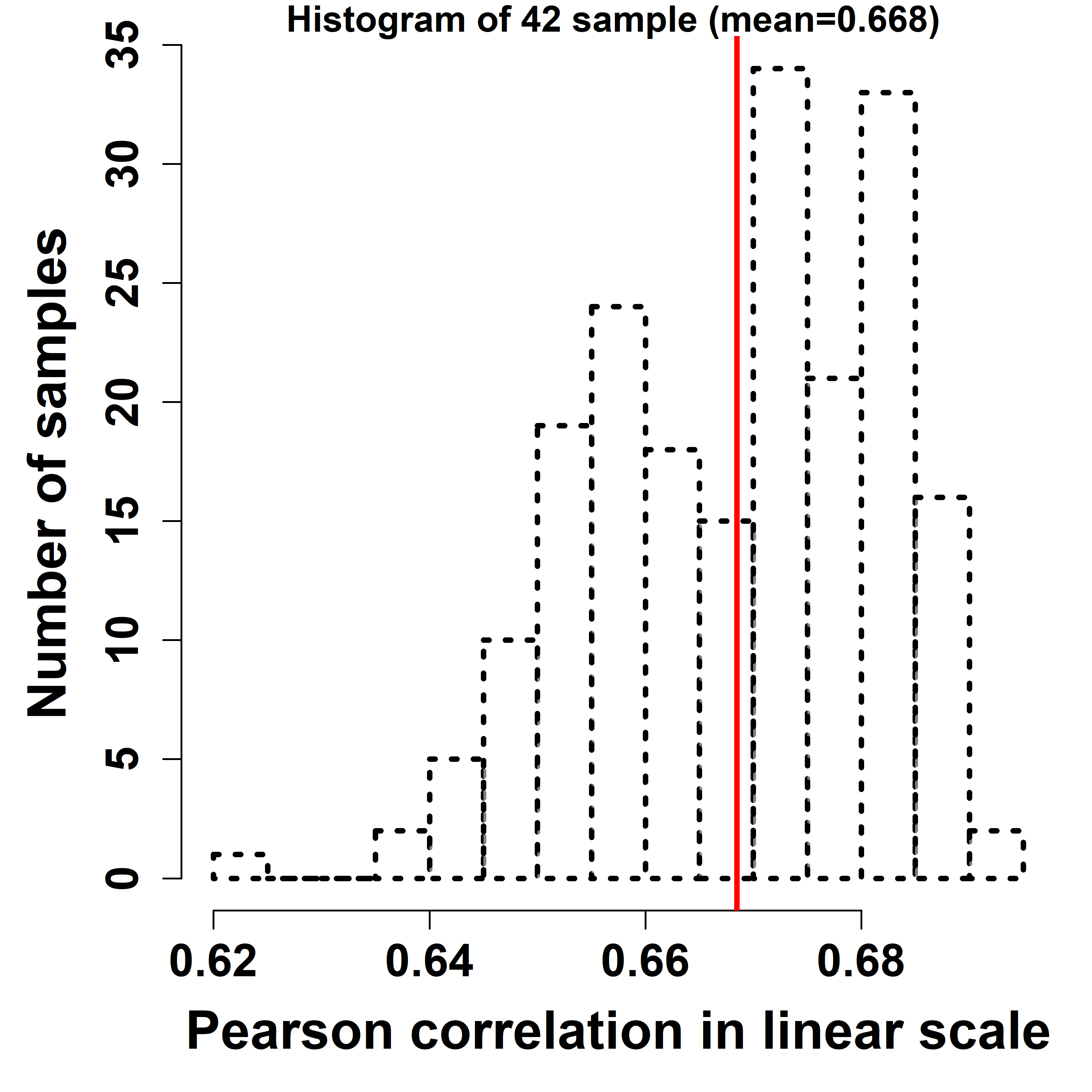}
    
    \includegraphics[width=0.31\linewidth]{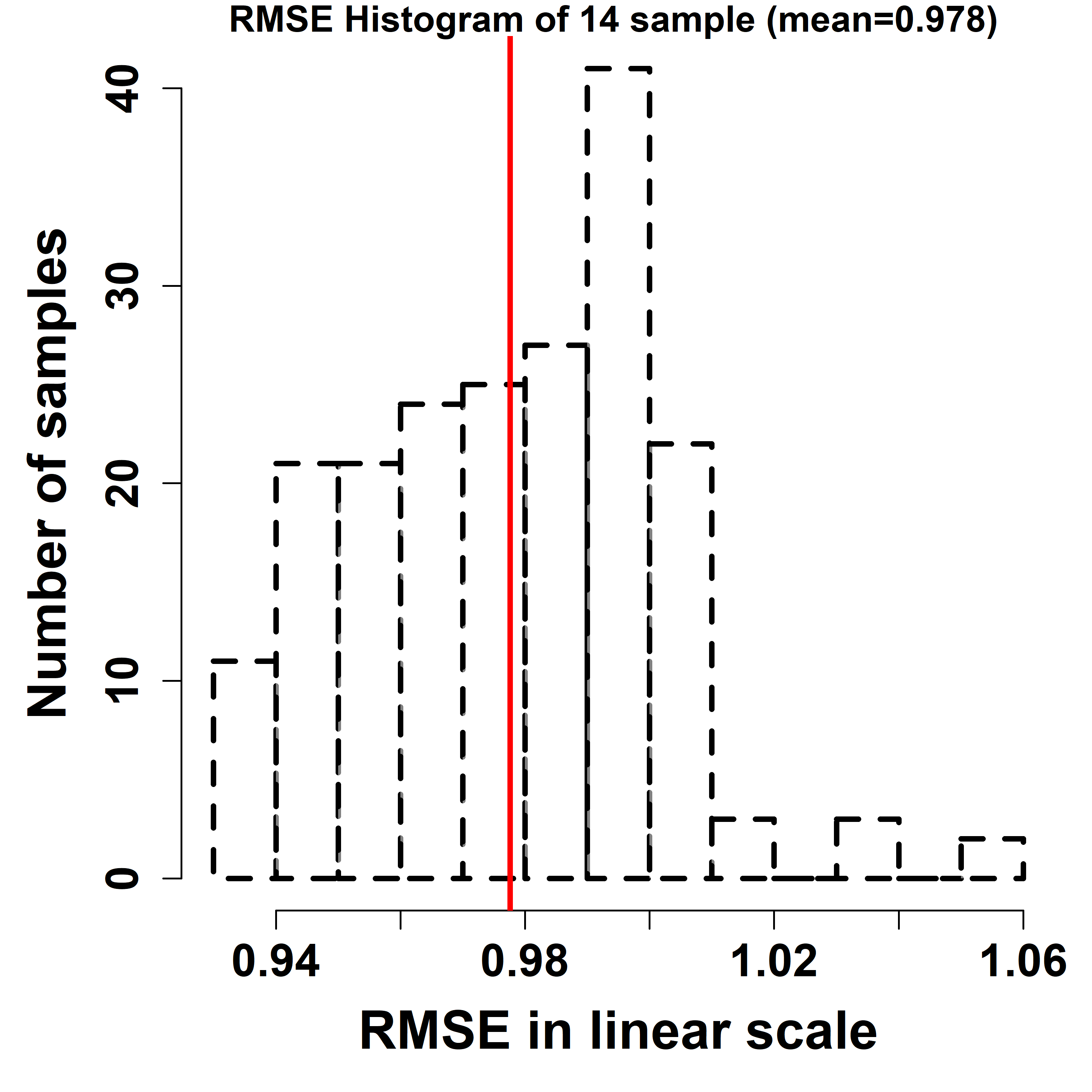}
    \includegraphics[width=0.31\linewidth]{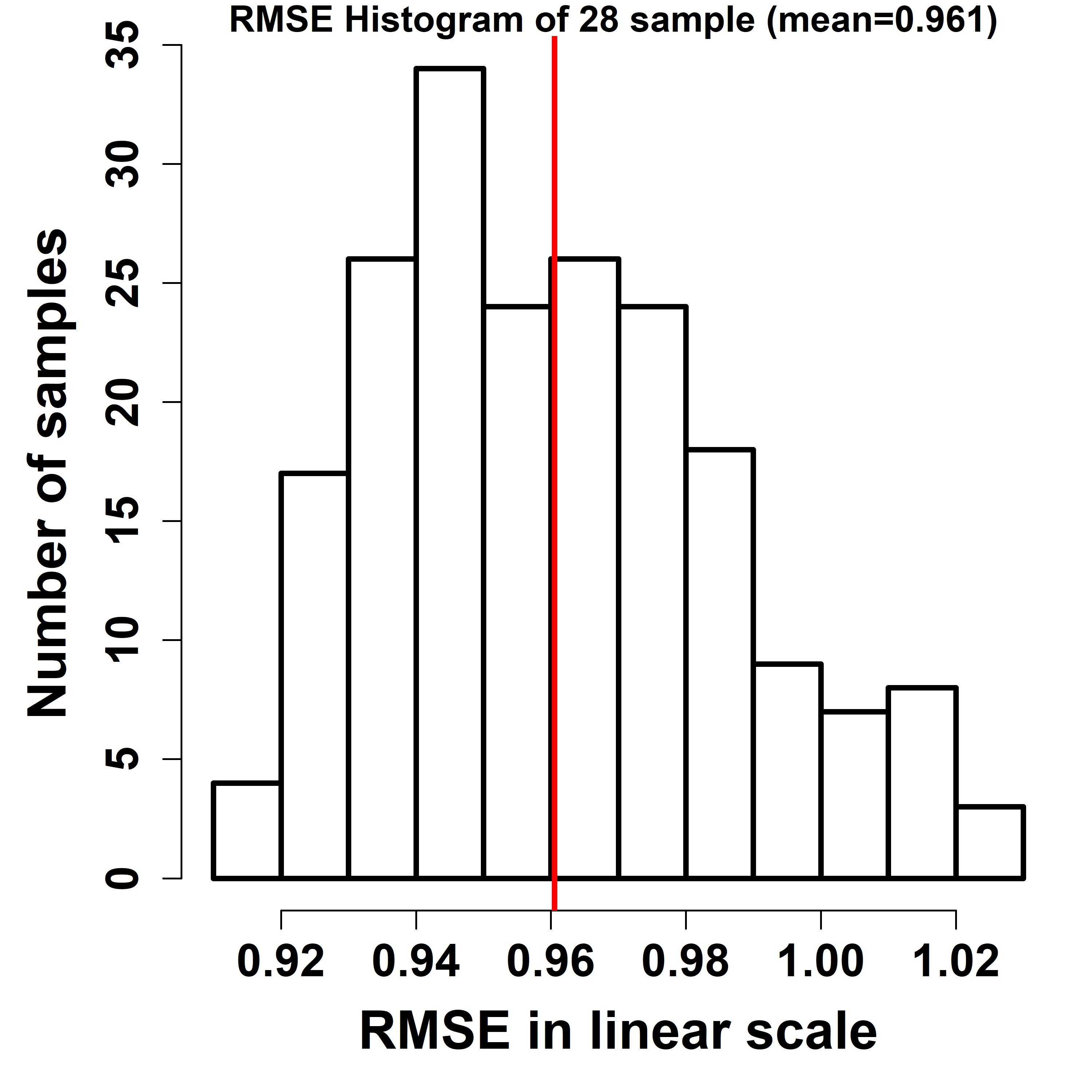}
    \includegraphics[width=0.31\linewidth]{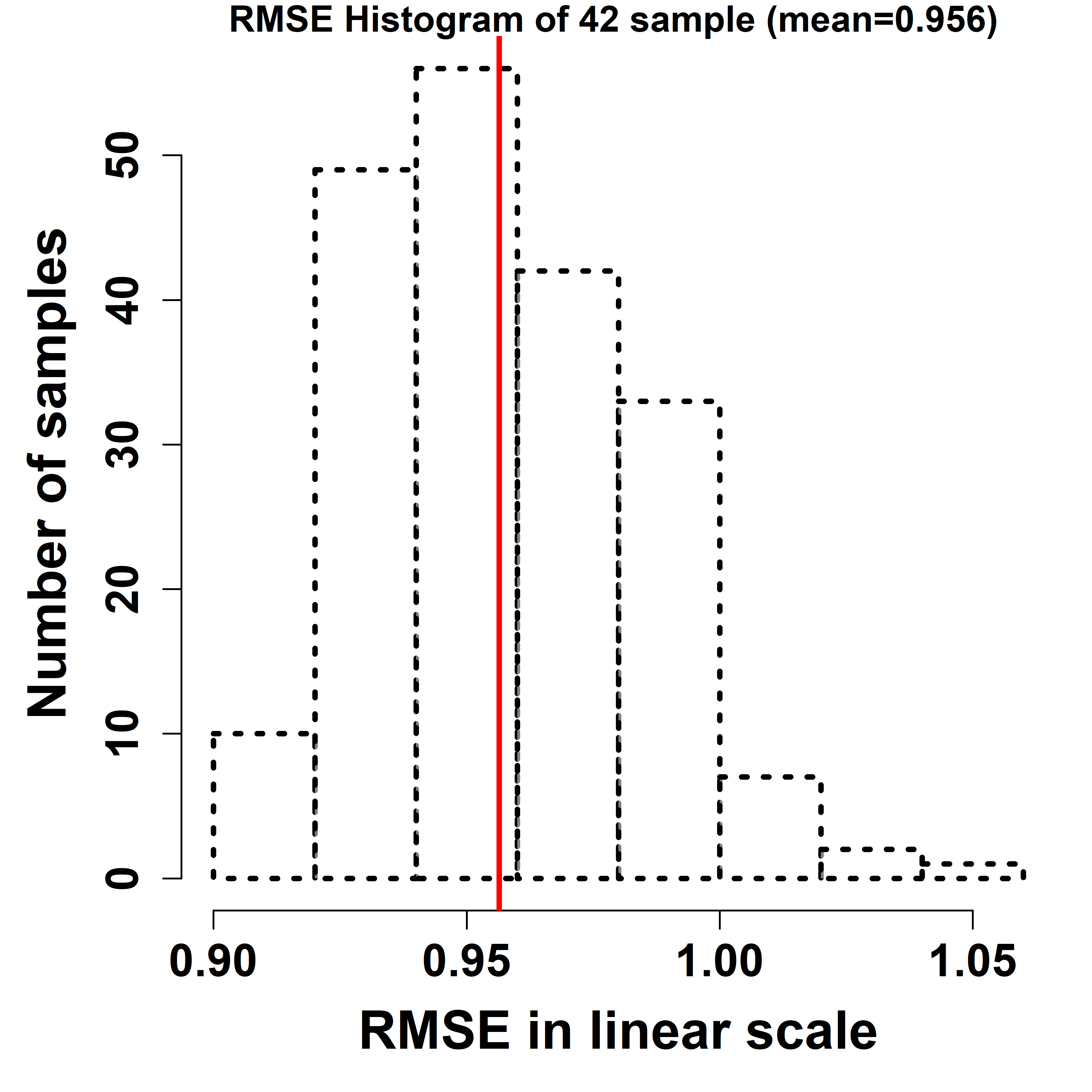}
    
    \includegraphics[width=0.31\linewidth]{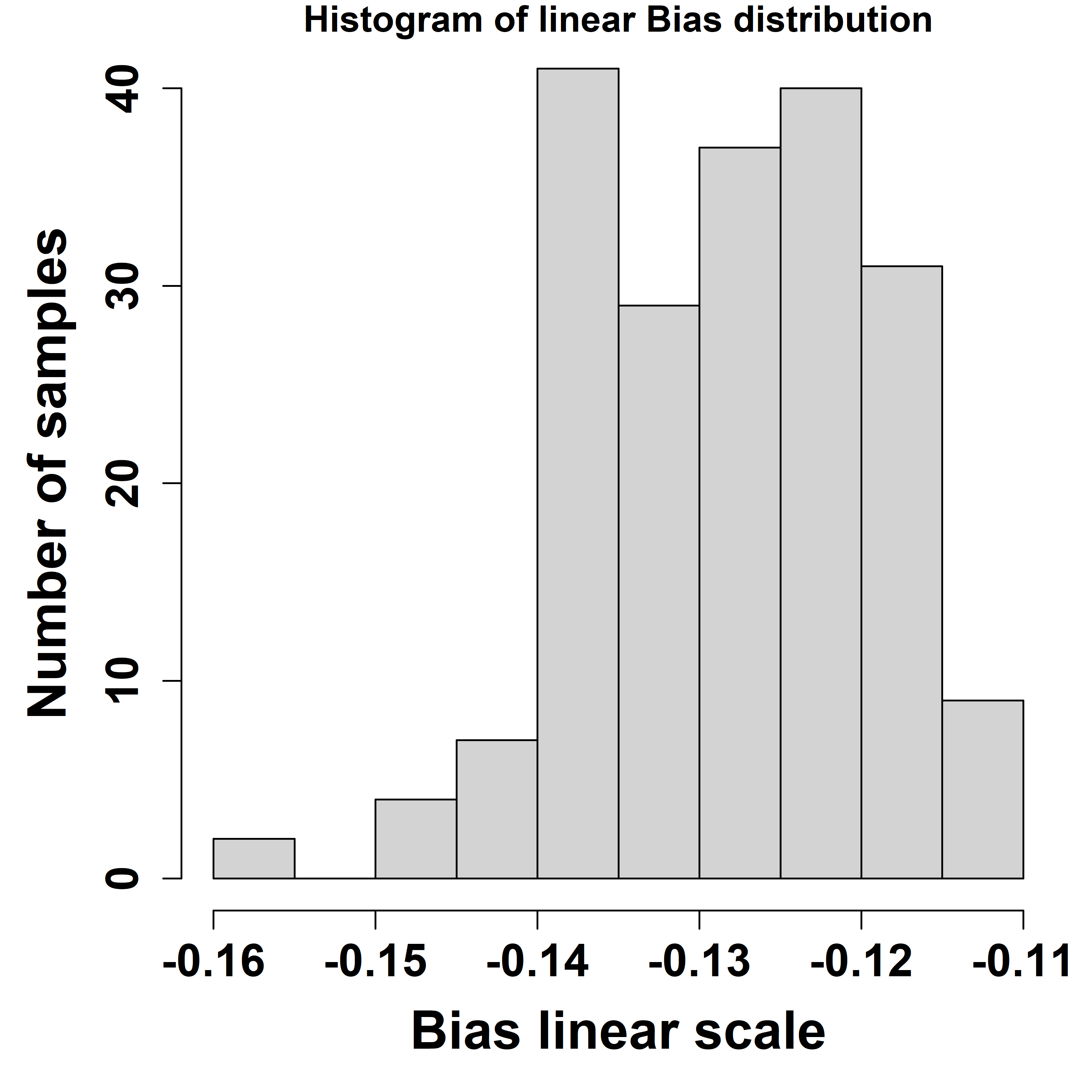}
    \includegraphics[width=0.31\linewidth]{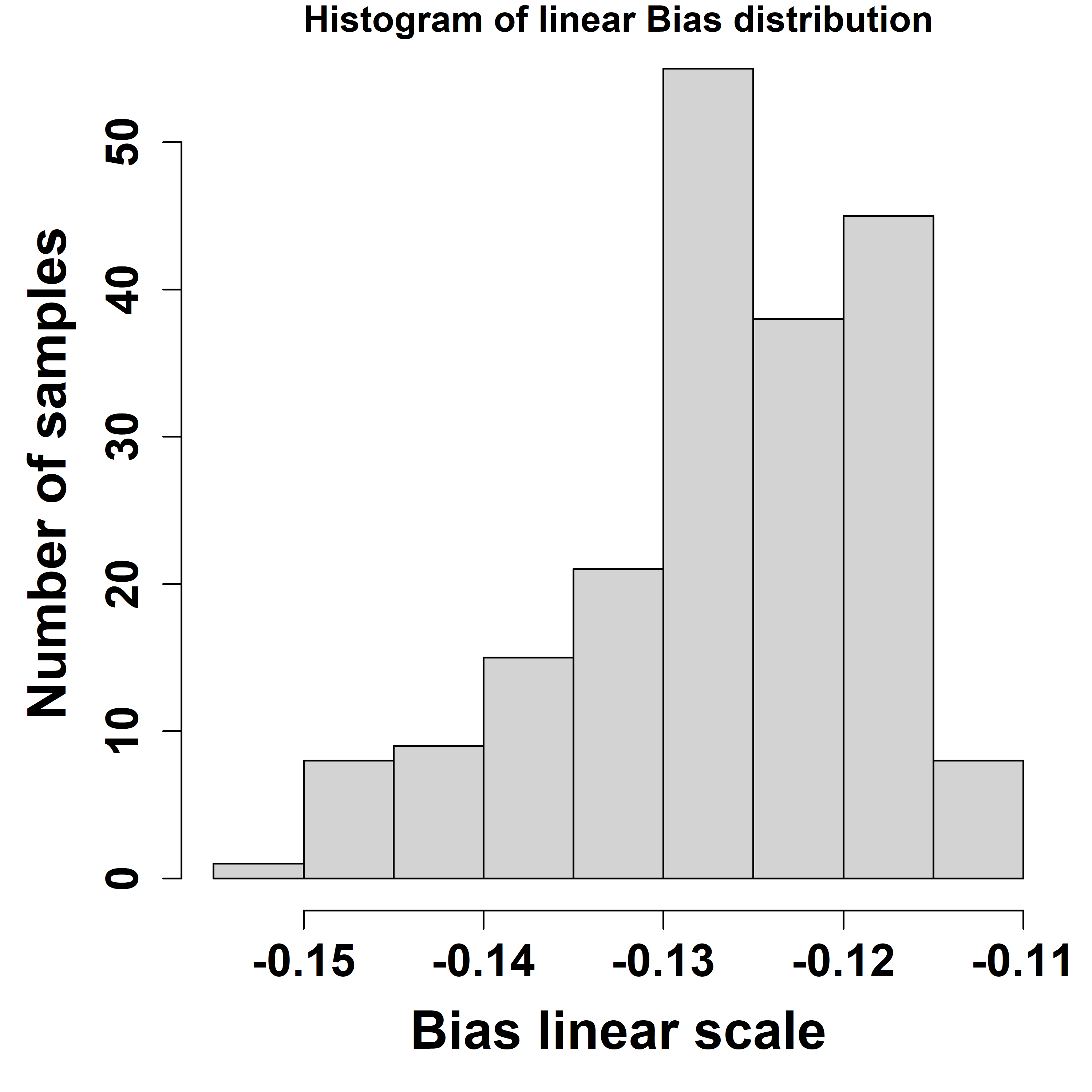}
    \includegraphics[width=0.31\linewidth]{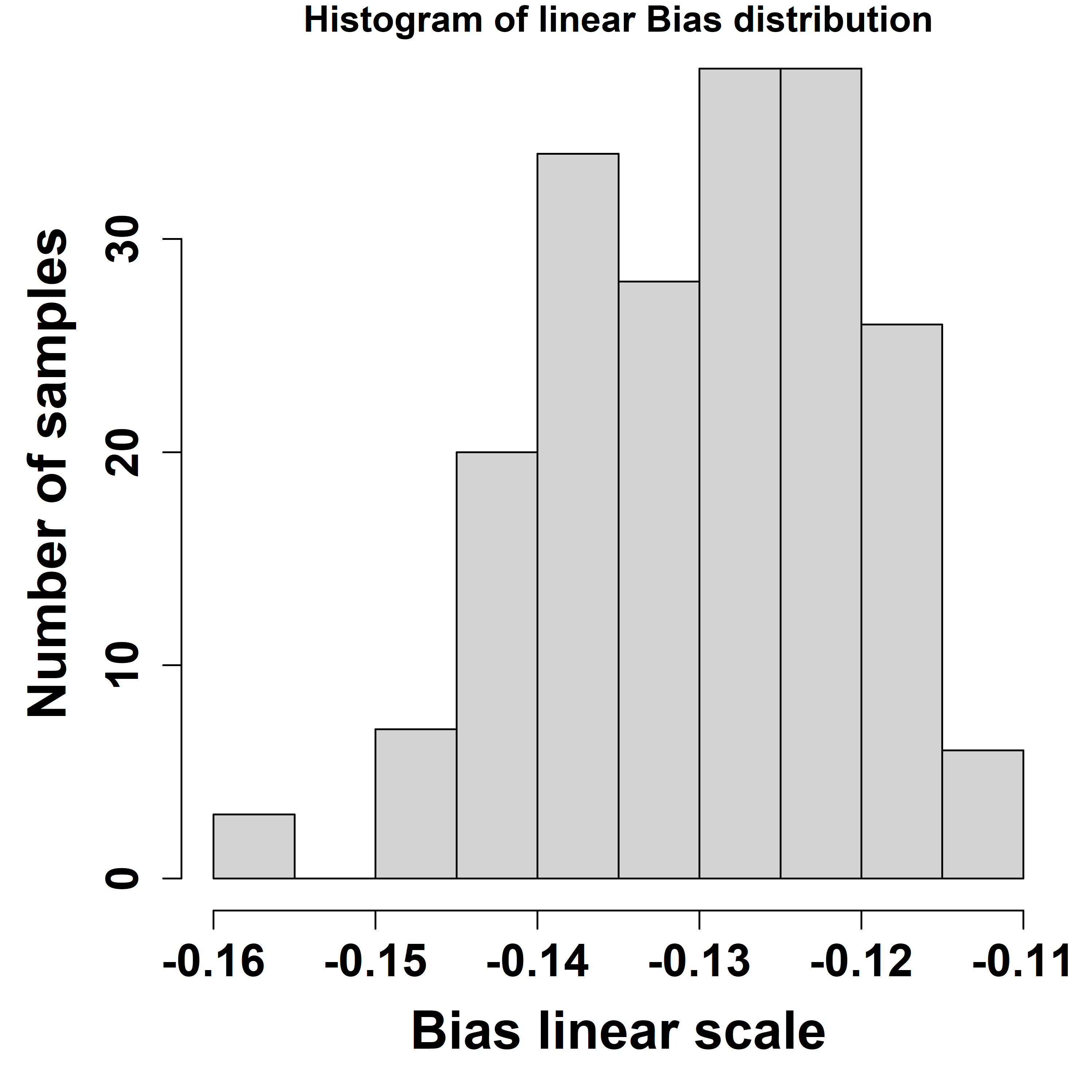}
    
    \includegraphics[width=0.31\linewidth]{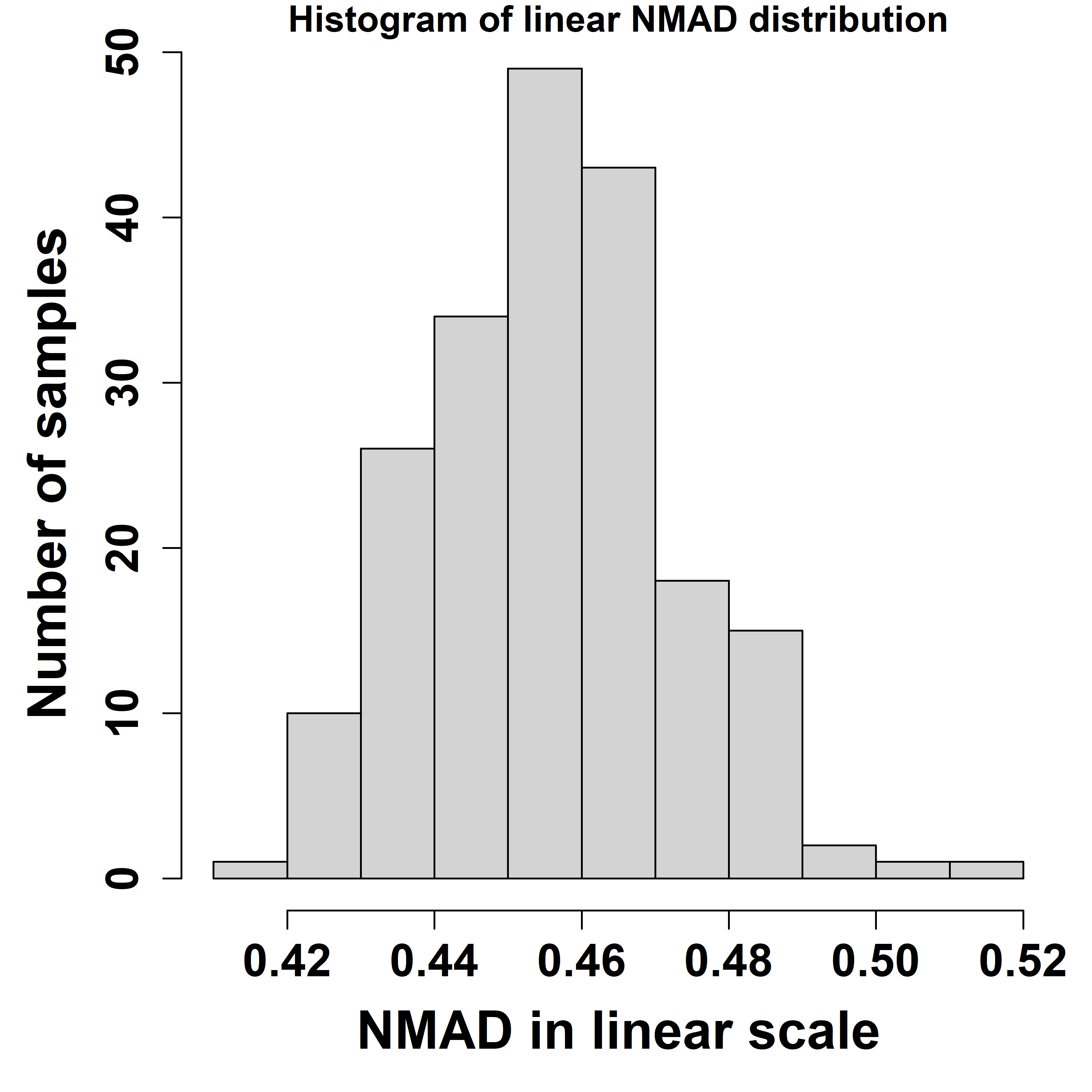}
    \includegraphics[width=0.31\linewidth]{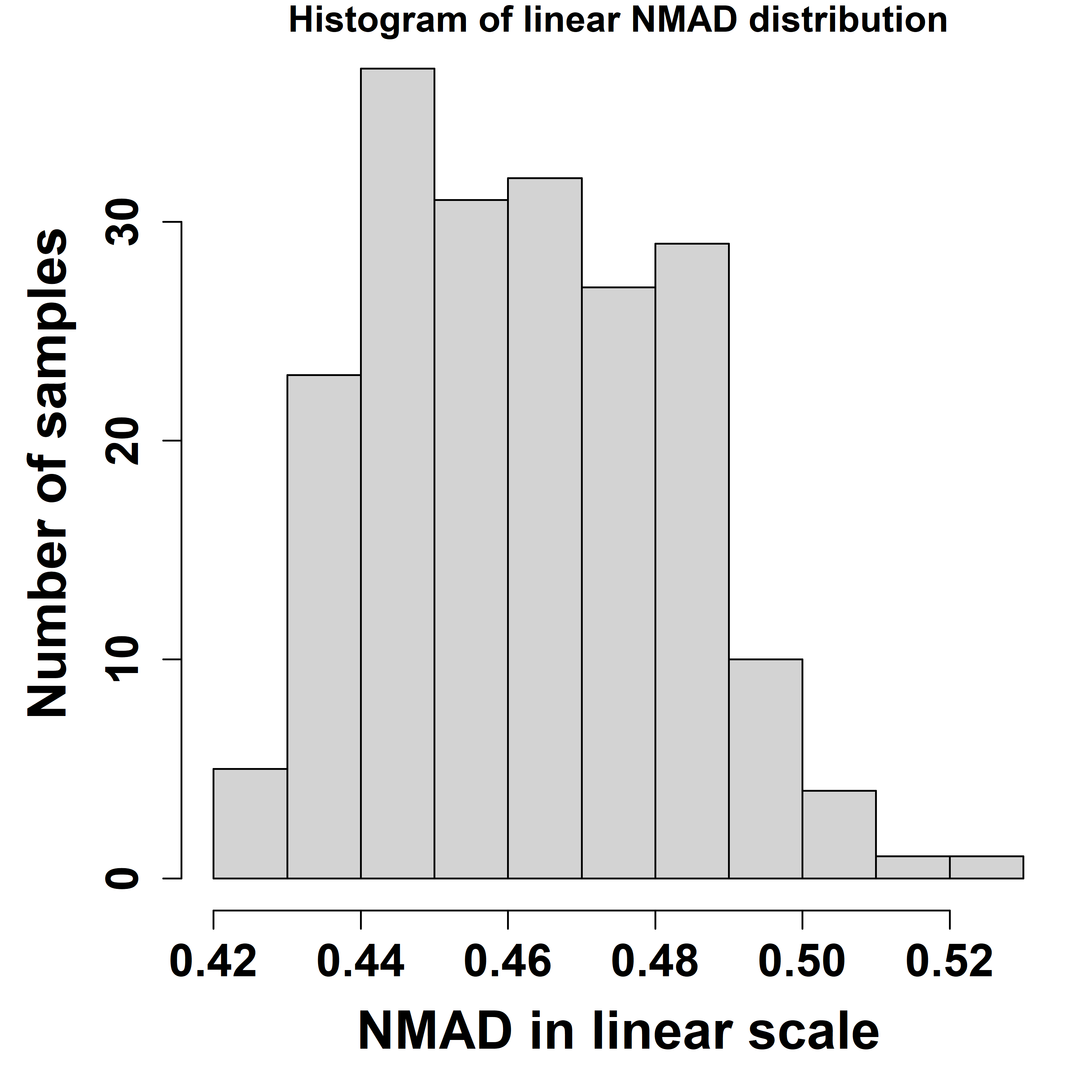}
    \includegraphics[width=0.31\linewidth]{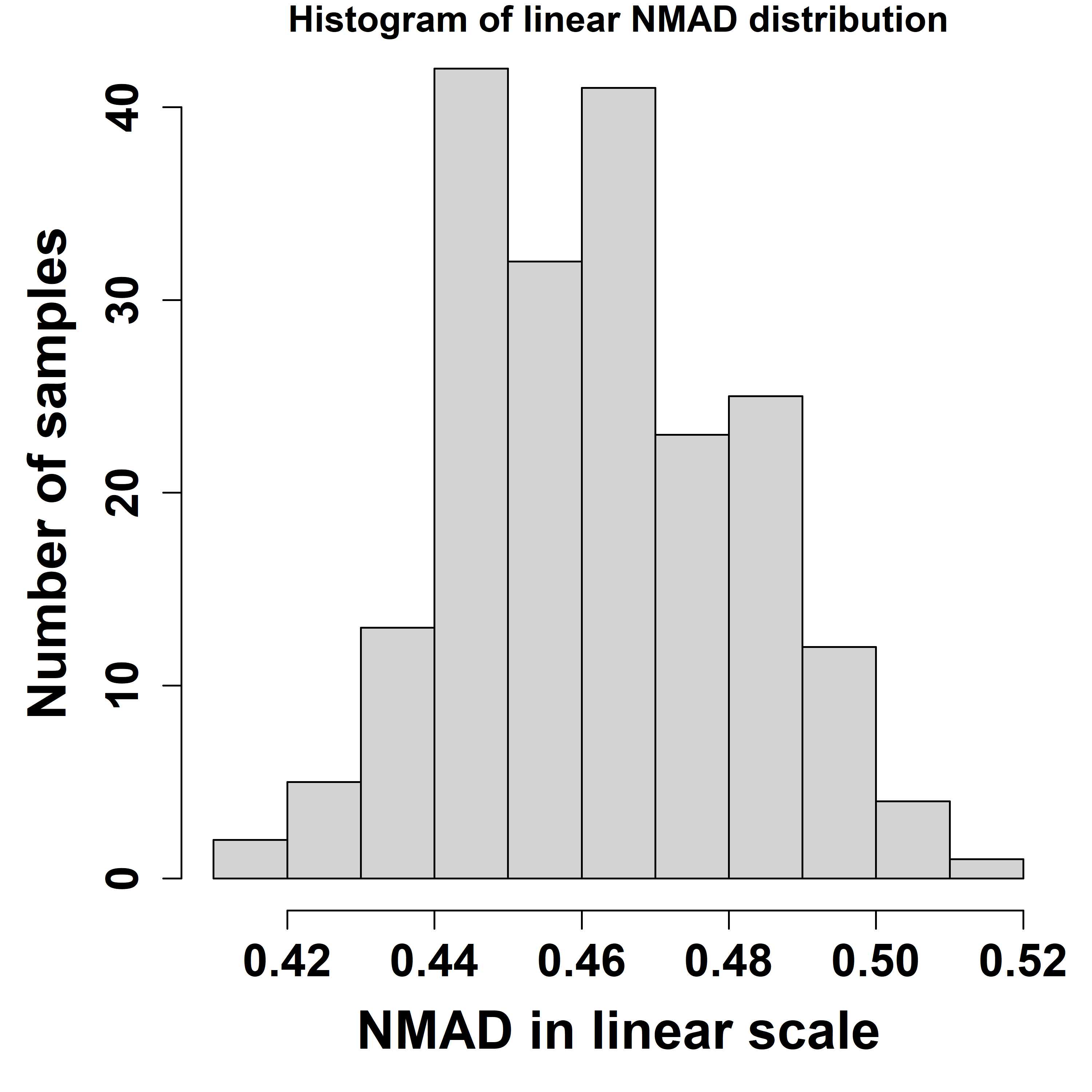}
   
    \caption{{\textit{Results of our MC analysis. All the plots shown are combined results of 200 simulated samples. 
   The first column of histograms shows the distribution of properties for the 14 sample cases. The second and third columns of histograms present the distribution of properties for 28 and 42 sample cases, respectively.
   The first row shows the correlation distribution, and the red line shows the mean. The second row shows the RMSE distribution. The third and fourth rows show the bias and NMAD distributions, respectively.}}
   }
    \label{fig:mcmc_simulation_results}
\end{figure*}

To understand if the result of an MC simulation is due to one realization only, we have then repeated the simulations with 200 different samples to de-randomize the process.
The MC simulation results shown in Fig. \ref{fig:mcmc_simulation_results} combine 200 instances of 14, 28, and 42 samples.
{Following the 10fCV analysis of each sample, we plot the $r$, RMSE, bias, and NMAD distributions for each sample size in columns 1, 2, and 3 for the samples of 14, 28, and 42, respectively.
The red vertical line in the $r$ and RMSE distributions shows the mean of the distribution.
}

{The 14-sample case: 
The top image shows the $r$ distribution. 
Here, we see the mean value of the distribution is 1.5\% higher at $r$=0.656 than the $r$ value of our SuperLearner results (see the upper left plot of Fig. \ref{fig:superlearner_results}).
In fact, 80\% of the simulated samples attained a $r$ greater than our SuperLearner result of $r$=0.646.
The mean value of RMSE is 3\% lower at RMSE=0.978 than our SuperLearner result.
}

{The 28-sample case: 
The mean values of the $r$ and RMSE distributions are 2.3\% higher at $r$=0.661 and 3.7\% lower at RMSE=0.961, respectively, than our SuperLearner results.
Here, 94\% of the simulated samples attained a $r$ greater than our SuperLearner result.
}

{The 42-sample case: 
Here, we see that the mean values of the $r$ and RMSE distributions are 3.4\% higher at $r$=0.668 and 4.4\% lower at RMSE=0.956, respectively, than our SuperLearner results.
Also, 97.5\% of the simulated samples attained a $r$ greater than our SuperLearner result.
}

{Based on the above-mentioned results, we can say that for the majority of simulated samples, there is an improvement in the results across all metrics.
This result is very promising because it indicates that our ML model generalizes to the GRB sample well and should give improved results for at least three years.
}

\subsection{Future of multi-wavelength redshift estimates via machine learning}\label{z_estimates}

The following subsection draws upon the results from Sec. \ref{mcmcresults} and Sec. \ref{generalization_results} for estimates on the number of GRB redshifts that can be predicted using ML analysis for future X-ray and optical observations of afterglows.
These estimates are based on the estimates presented in \citep{Dainotti2022b}, which in turn rely on the existing X-ray and optical samples determined by \cite{Dainotti2024GRBRedshift} and \cite{Dainotti2024ApJ...967L..30D}, provided up to May 2025.

\begin{table*}[!ht]
    \centering
    \caption{Present and future numbers of GRB observations.}
    \begin{tabular}{|m{3cm}|m{2cm}|m{3.25cm}|m{1.25cm}|m{1.25cm}|m{2.5cm}|m{2cm}|}
    \hline
        \textbf{GRB subsets} & \textbf{Number} & \textbf{Current data gathered from-to} & \textbf{Number of yrs} & \textbf{Rate/yr} & \textbf{Data gathering until September 2025} & \textbf{GRBs by September 2025} \\ \hline\hline
        Optical Afterglow with z  & 536 & 05/1997 - 10/2023 & 27 & 19.7 & 10/2023-09/2025 & 574 \\ \hline
        Optical plateau with z  & 180 & 05/1997-05/2021 & 24 & 7.50 & 05/2021-09/2025 & 193 \\ \hline
        Optical Afterglow without z & 310 & 05/1997- 10/2023 & 26 & 11.92 & 10/2023-09/2025 & 333 \\ \hline
        Optical plateau without z  & 104 & 05/1997- 10/2023 & 26 & 4.00 & 10/2023-09/2025 & 112 \\ \hline
        X-ray plateaus without z & 299 & 01/2005 - 02/2024 & 19.1 & 15.7 & 02/2024-09/2025 & 324 \\ \hline
        X-ray plateaus with z & 255 & 03/2005 - 12/2023 & 18.7 & 13.6 & 12/2023-09/2025 & 279 \\ \hline\hline
        Known redshifts in X-rays \& Optical by May 2025 & 279 (X-ray) + 193(OA) & From May 1997 to September 2025 & ~ & ~ & ~ & 472 \\ \hline
        Predicted redshift via ML in X-rays and Optical by May 2025 & 324 (X-ray) + 112 (OA) & ~ & ~ & ~ & ~ & 436 \\ \hline
        \% increase from X-rays and Optical plateaus with z with ML & 436/472 & ~ & ~ & ~ & ~ & \textbf{92\% more redshifts will be estimated} \\ \hline
    
    \end{tabular}
    \tablefoot{\textit{Number of GRBs this study will enable to have redshift predictions in the future using Swift observations.}}
    \label{fig:grb_table}
\end{table*}

%\begin{table*}
%\caption{Present and future numbers of GRB observations.}
%    \centering
%    \includegraphics[width=\textwidth]{Figures/grb_table_110924.png}
%    \tablefoot{\textit{Number of GRBs this study will enable to have redshift predictions in the future using Swift observations.}}
%    \label{fig:grb_table}
%\end{table*}

For the X-ray afterglow, we currently have 255 GRBs possessing plateaus and measured $z_{obs}$, observed from 03/2005 to 12/2023.
We have 299 events with X-ray plateaus but without redshifts, observed from 01/2005 to 02/2024. 
Both sets of GRBs are taken from the Swift/XRT catalogs.
Based on these data, $\sim$13.6 GRBs with X-ray plateaus and $z_{obs}$, and $\sim$15.7 GRBs with X-ray plateaus and unknown redshifts are expected annually.
For GRBs with $z_{obs}$ and X-ray plateau, 24 additional GRBs are expected from 12/2023 until 09/2025.
For GRBs with X-ray plateaus and without $z_{obs}$, 25 additional GRBs are also expected from 02/2024 until 09/2025.
This sample with unknown redshift will be the generalization set.
So, the number of GRBs with $z_{pred}$ from ML analysis will be greater than 324. 

For the optical afterglow, we have gathered 536 LCs through the 416 ground-based Telescopes, the HST, the Swift/UVOT satellites, the GCN, and private communications.
These 536 GRBs, observed from 05/1997 to 10/2023, have $z_{obs}$ and optical afterglow. 
Out of these 536 GRBs, 180 (34\%) GRBs have optical plateau observations.
As of this writing, according to the Greiner webpage\footnote{https://www.mpe.mpg.de/~jcg/grbgen.html}, the total number of GRBs with optical afterglow is {932}, and {622} have secure redshifts.
Thus, 310 GRBs (932 - 622) do not have redshift but have an optical afterglow.
Assuming the same fraction of plateau observations in the optical afterglow, our current generalization set would be 34\% of 310, which is 104 GRBs.
If we extrapolate to 09/2025, we expect to have 112 GRBs with an optical plateau and no measured redshifts.
Similarly, we expect to have 193 GRBs by 09/2025, which have an optical plateau and $z_{obs}$.
Thus, we expect more than 112 GRBs with $z_{pred}$ from ML analysis. Thus, the new $z_{pred}$ will be more than {436 (324+112), where 324} and 112 are GRBs with X-ray and optical plateaus, respectively.
With this increase of 92\% in the redshift sample, we will enhance the accuracy of the LF and density rate evolution estimates, and thus it will help address the discrepancy between the GRBFR and SFR evolution.
The above numbers are also reflected in the Table \ref{fig:grb_table}.
{This analysis can be extended for future reference, and it serves as our foundation to compute how many GRBs we will approximately have in the next one, two, and three years from now.}

\cite{Dainotti2022MNRAS.514.1828D} also calculated the number of GRB observations required to achieve similar precision on $\Omega_M$ for the flat $\Lambda$CDM model as obtained in studies by \cite{Conley2011,Betoule2014,Scolnic2018} (for SNe Ia). 

According to Table.9 of \cite{Dainotti2022MNRAS.514.1828D}, to reach the precision of \cite{Conley2011}, the lowest number of X-ray platinum sample GRBs with redshift needed, with half error bars on plateau parameters, is 354. 
To achieve this number, we assume that by using an updated LC reconstruction (LCR) technique, an improvement of the methodology from \cite{Dainotti2023a}, which is in preparation, we will have the same number of GRBs with half error bars (denoted by n=2) as with full error bars (denoted by n=1).
Currently, we have 105 GRBs that belong to the platinum sample with $z_{obs}$.
Assuming the same fraction of platinum GRBs exists in the generalization set, we should have another 125 GRBs from the generalization set used in this work.
Combining these two, we estimate that we currently have 230 platinum sample GRBs with pseudo-redshifts. 
Further, considering the observations we will obtain from SVOM (\cite{ECLAIRS_Godet_2014} and SVOM webpage\footnote{https://www.eoportal.org/satellite-missions/svom-eclairs-gamma-ray-imager}) and the application of ML and LCR, we can expect to reach the precision of \cite{Conley2011} by 2029 (see PLATtrim20 column of Table.\ref{tab:xray_estimates}).
Thus, this ML effort reduces the number of years required to reach the precision of \cite{Conley2011} by 26 years.
Similarly, to reach the precision of \cite{Betoule2014}, we need the observational capabilities of THESEUS.
According to our estimates in Table.\ref{tab:xray_estimates}, again, using LCR and ML to boost the numbers and GRBs with half error, we should be able to reach the precision of \cite{Betoule2014} by 2041 at least.

Similarly, let us consider the optical estimates from Table 9 of \cite{Dainotti2022MNRAS.514.1828D}. 
We should be able to reach the precision of \cite{Conley2011} with the current data we have if we assume that we will have an equal number of half error GRBs as we have full error GRBs, achieved by using updated LCR methodology. 
If we, however, trim the sample in the closest 25 to the 3D fundamental plane (OPTtrim25 column of Table.\ref{tab:opt_estimates}), we would have reached the precision of \cite{Conley2011} even in 2005, 6 years earlier than the one reached by the SNe Ia.
We note that the numbers mentioned along the fourth row of Table.\ref{tab:opt_estimates} are estimated based on GRBs observed from January 2005. 
This is done to avoid negative numbers, as some of the estimates are reached well before 2024. 
The corresponding year estimates are obtained based on the rate of observation using Swift and other ground-based optical observatories.
Furthermore, if we take into account the rate of observation of SVOM and the fraction of optical GRBs with the plateau we currently have, then we can reach the precision of \cite{Conley2011} by 2028 (only 4 years from now), with the current error bars.
However, to reach the precision of \cite{Betoule2014}, we need to consider the observation of THESEUS as well.
Considering THESEUS's rate of observation to be 700 GRBs per year \citep{Dainotti2022MNRAS.514.1828D,Amati2018AdSpR..62..191A}, we again consider the same fraction of optical plateau GRBs as we currently have.
With this, we can reach the precision of \cite{Betoule2014} by 2039, 7 years after the launch of THESEUS (the current expected launch is 2032, if the mission is approved).
However, to reach the same precision as \citep{Scolnic2018}, we would need to wait at least until 2041 and use both ML and LCR and half error bars.

However, we can say that these estimates are rather conservative since new optical telescopes and facilities will be operating in the future, and we can rely on the efforts of some of us to continue building the largest optical GRB catalog to date \citep{OpticalCatalog2024MNRAS.533.4023D}. 
Indeed, one of the outcomes of gathering multiple data points from several sources has the advantage of allowing an increase of data points, in general, in the LCs and, in particular, in the plateau region when it occurs. 
Thus, we assumed that once the optical GRB sample is updated, our estimates will be larger in terms of the total number of optical plateaus, and therefore the number of years needed to reach SNe Ia precision will be less. 
In addition, regarding the X-ray sample, some of us are currently working on choosing a subsample, not driven by phenomenological reasons but driven by the magnetar model \citep{Zhang2001ApJ...552L..35Z,Rowlinson2014MNRAS.443.1779R,Yi2017JHEAp..13....1Y}. 
In such a case, we can have a smaller scatter of the plane as shown for the 2D Dainotti relation between the X-ray luminosity at the end of the plateau and its rest-frame duration \cite{Dainotti2008,Dainotti2011,Dainotti2013,Dainotti2017,Wang:2021hcx}, and we can reach again the precision of the SNe Ia 
sooner, using the X-ray sample.

\begin{table*}[h]
\caption{Forecast on the future precision of X-ray GRB cosmology.}
    \centering    
    \begin{tabular}{@{}lccc@{}}
        \toprule
        Sample & PLAT & PLATtrim10 & PLATtrim20 \\ 
        \midrule
        \midrule
        \textbf{Reaching Precision of Conley (before launch of THESEUS)} \\
        \midrule
        Original estimates (n=1) & 789 & 847 & 646 \\
        \midrule
        n=2 + LCR + ML & 127 & 169 & 124 \\
        Years to reach (n=2 + LCR + ML) & 2029 & 2031 & 2029 \\
        \midrule
        \midrule
        \textbf{Reaching Precision of Betoule (after launch of THESEUS)} \\
        \midrule
        \midrule
        Original estimates (n=2) & 1452 & 1788 & 1466 \\
        \midrule
        n=2 + LCR + ML & 1040 & 1376 & 1054 \\
        Years to reach (n=2 + LCR + ML) & 2041 & 2044 & 2041 \\
        \bottomrule
    \end{tabular}
    \tablefoot{\textit{Number of platinum sample X-ray GRBs with redshift needed to reach the precision of $\Omega_M$ as obtained in studies by \cite{Conley2011,Betoule2014,Scolnic2018} (for SNe Ia). This is an extension of the table provided in \cite{Dainotti2022MNRAS.514.1828D}.
    The term n=1 indicates that the error bars on the GRB plateau parameters are full width, while n=2 indicates that the error bars are halved.
    The "LCR" indicates that the estimates for the number of GRBs have been boosted by using LCR, while "ML" indicates that the estimates have been boosted by using ML.}}
    \label{tab:xray_estimates}
\end{table*}

\begin{table*}
\caption{Forecast on the future precision of optical GRB cosmology.}
    \centering    
    \begin{tabular}{@{}lccc@{}}
        \toprule
        Sample & OPT & OPTtrim10 & OPTtrim25 \\ 
        \midrule
        \midrule
        \textbf{Reaching Precision of Conley (before launch of THESEUS)} \\
        \midrule
        Original estimates (n=1) & 271 & 330 & 244 \\
        \midrule
        n=1 + LCR + ML & 99 & 158 & 72 \\
        Years to reach (n=1 + LCR + ML) & 2032 & 2037 & 2030 \\
        \midrule
        Original estimates (n=2) & 142 & 112 & 36 \\
        \midrule
        n=2 + LCR* & 30 & 79 & 3 \\
        Years to reach (n=2 + LCR) & 2028 & 2024 & 2005 \\
        \midrule
        \midrule
        \textbf{Reaching Precision of Betoule (after launch of THESEUS)} \\
        \midrule
        Original estimates (n=1) & 1031 & 829 & 685 \\
        \midrule
        n=1 + LCR + ML & 770 & 568 & 424 \\
        Years to reach (n=1 + LCR + ML) & 2044 & 2041 & 2039 \\
        \midrule
        Original estimates (n=2) & 284 & 393 & 350 \\
        \midrule
        n=2 + LCR &  118 & 227 & 184 \\
        Years to reach (n=2 + LCR) & 2035 & 2037 & 2036 \\
        \midrule
        n=2 + LCR + ML & 23 & 132 & 89 \\
        Years to reach (n=2 + LCR + ML) & 2032 & 2034 & 2033 \\
        \midrule
        \midrule
        \textbf{Reaching Scolnic Precision (after launch of THESEUS)} \\
        \midrule
        Original estimates (n=2) & 1086 & 1513 & 822 \\
        \midrule
        n=2 + LCR + ML & 825 & 1252 & 561 \\
        Years to reach (n=2 + LCR + ML) & 2045 & 2052 & 2041 \\
        \bottomrule
    \end{tabular}
    \tablefoot{\textit{Number of optical GRBs with a plateau and redshift needed to reach the precision of $\Omega_M$ as obtained in studies by \cite{Conley2011,Betoule2014,Scolnic2018} (for SNe Ia). This is an extension of the table provided in \cite{Dainotti2022MNRAS.514.1828D}.
    The term n=1 indicates that the error bars on the GRB plateau parameters are full width, while n=2 indicates that the error bars are halved.
    The "LCR" indicates that the estimates for the number of GRBs have been boosted by using LCR, while "ML" indicates that the estimates have been boosted by using ML.}
    }
    \label{tab:opt_estimates}
\end{table*}

\subsection{Applicability to future missions}\label{future}

{It is important to note that all the methodologies spelled out here can be applied not only to GRBs observed by Swift, but also to any GRB for which we can observe the plateau emission. 
Thus, we envision this method can be applied to GRBs observed by {Space-based multi-band astronomical Variable Objects Monitor (SVOM,\cite{SVOMwei2016deeptransientuniversesvom})}, the Transient High-Energy Sky and Early Universe Surveyor (THESEUS) mission  \citep{Amati2018AdSpR..62..191A}, and {the Einstein Probe mission}\citep{EinsteinProbeYuan:2015y8}. 
Surely, with the new data from other satellites, one needs to be cautious about how the variables' parameter spaces will change compared to the space of the Swift variables and how much this will impact the prediction.
{Properly leveraging the power of these new satellites will require a re-training of the ML models. However, the base methodology described here and in our previous works \citep{Dainotti2024GRBRedshift,Dainotti2024ApJ...967L..30D} should still produce reliable results.}
}

{The SVOM mission, in particular, is very promising indeed, as one of their key aims is to quickly identify and observe the afterglow of GRBs at both X-ray and visible wavelength. The SVOM mission was launched in June 2024.
The methodology described here requires the GRB to have a plateau emission. Thus, SVOM's ability to quickly slew to a GRB's location will enable quick follow-up of afterglow emission.
This will lead to more observations of the plateau, especially at X-ray wavelengths, and make the GRBs observed by SVOM compatible with our methodology, albeit with careful consideration of their parameter spaces.
}

{According to \cite{SVOM2018MmSAI..89..266C}, the planned mission duration for SVOM is 3 years. 
Based on estimates by \cite{ECLAIRS_Godet_2014}, 200 GRBs are expected to be observed by SVOM over its 3-year lifetime.
However, considering that satellites such as Swift, Chandra, INTEGRAL, and Hubble have lasted well beyond their originally intended lifetime, we can safely assume that SVOM will continue to work beyond its 3-year lifetime.
Thus, based on estimates from \cite{Dainotti2022MNRAS.514.1828D}, we take 15 years (up to 2039) as the estimated lifetime for SVOM. After this, the observational duties would be taken up by the THESEUS mission.
Following these assumptions, we estimate that SVOM will make around 1005 GRB observations over its extended lifetime.
Out of these 1005 GRBs, we expect 825 GRBs to be detected in X-ray. 
This is derived from the fraction of GRBs detected in X-ray by Swift. 
Next, based on the data presented here, we can estimate that around 38\% (318 GRBs) of these GRBs will exhibit the plateau feature.
Furthermore, of these 318 GRBs with plateau, we expect 46\% (147) to have spectroscopic redshift measurements. Again, this estimate is based on current Swift data.
If we train an ML model on these 147 GRBs, we should be able to predict the redshift of the 171 GRBs (318 - 147). 
This will effectively increase the number of GRBs from SVOM that can have pseudo-redshifts by 117\%.
We note that we have not considered the degradation of the SVOM's orbit over time.
However, these estimates are a lower limit on SVOM's actual observation rates, as we expect it to be more efficient.}
{THESEUS is expected to identify around 20 events at $z > 6$ in 3.5 years of nominal operation; hence, it will  provide a much larger sample of high redshift GRBs than achieved so far \citep{Tanvir2021ExA....52..219T}. 
High$-z$ Gundam mission might be comparable to THESEUS (five events per year, \cite{Yonetoku2023HEAD...2010318Y}). 
SVOM is already operational and equipped with the 4 – 150 keV wide-field trigger camera and is sensitive to a wide variety of GRBs, including those at high redshifts. It is expected to detect around 80-100 GRBs per year.}

\subsection{Redshift inference with the web app}

\begin{figure}[!h]
    \centering
    \fbox{\includegraphics[width=0.475\textwidth]{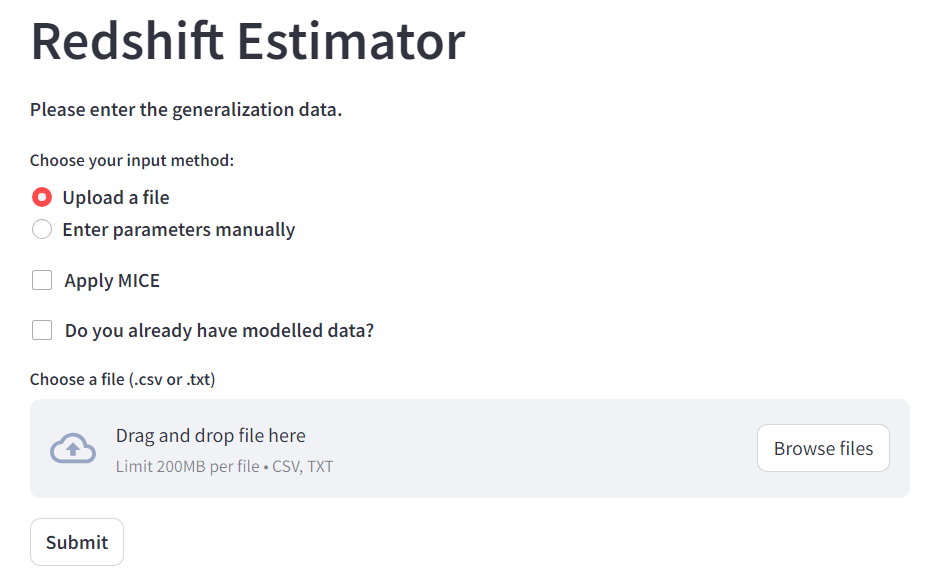}}
    \caption{\textit{Redshift Estimator, the homepage of the app, depicted with the options of choosing MICE, uploading a file, entering parameters manually, and using custom models.}}
    \label{fig: App_Layout}
\end{figure}

{Our web app has various functionalities, but its main focus is inferring the redshift of GRBs without $z_{obs}$. 
Users have the flexibility to upload a file containing the predictors discussed in this work (see Sec. \ref{lasso}) or manually enter the parameters of a single GRB for redshift estimation.
This interface is presented in Fig. \ref{fig: App_Layout}.
This module of the web app also allows the user to obtain MICE imputations if their data contains missing values. To do so, they select the `Apply MICE' button.
After providing the data, the web app initiates the back-end ML codes, the results and performance of which have been presented in this work.
Upon successful execution, the user is presented with individual redshift values for each GRB {in a tabular manner, which can also be downloaded as a text file.}
This table displays the GRB IDs along with $z_{pred}$ and its associated uncertainties. 
{Furthermore, if a user already has their custom model, trained from modules such as Model Comparision and SuperLearner, then they can use that model for predictions. 
For this, they need to select the "Do you have modeled data?" option. 
However, the users can use the model trained on our data without necessarily providing their own training set.}
Finally, comprehensive documentation is available on the web app in the `Documentation' module to help users navigate. 
}

\section{Discussion}\label{discussion}

{The findings from \cite{Dainotti2024ApJ...967L..30D} demonstrate the promise of using machine learning for redshift estimation, particularly in overcoming the limited number of GRBs with precise redshift measurements. 
In this work, we have expanded both the methodology and the sample sizes for the training and generalization data sets.}
Providing redshift predictions for the generalization set will help future population studies address the GRBFR and SFR discrepancy and other relevant issues.
{In the future, the $z_{pred}$ values obtained here can also be utilized to explore the characteristics of the central engine in GRB events, focusing on GRBFR, LF, and the structure of LCs.}
{We can also use GRB LCs obtained from reconstruction, which some of us did in \cite{Dainotti2023a}.
We demonstrated the effectiveness of this reconstruction technique in reducing the errors on plateau parameters by an average of 30\%-40\%.
}

{
Another point we wish to highlight here is that even though we have the largest sample of LGRBs with plateau and redshift, this is still at least an order of magnitude less than the data required to reliably train more complex ML models, such as neural networks. 
Indeed, without including the plateau, Swift XRT observed 1420 GRBs up to December 2024, which is too small to train complex ML models properly.
In some of our previous works, using Active Galactic Nuclei data \citep{Dainotti2021ApJ...920..118D,Narendra2022,gibson2022}, we have shown that SuperLearner picks different, more complex models (e.g., XGBoost, EARTH) for the larger sample size.
Similarly, we tested multiple ML models to find the best set of models that can give accurate redshift predictions with the current data set. 
The three chosen models performed the best during our analysis. 
Given the size of our data set, parametric and semi-parametric functions such as GLM and GAM perform better.
This fact is further highlighted in the Appendix in Fig. \ref{model_weight}, where the weight assigned by SuperLearner to the model EARTH is lower than the weights assigned to the GAM, GLM, and RF models. 
This is in contrast with the results presented in \cite{Narendra2022} where Earth was one of the best models.
Furthermore, to test whether the performance of these models is maintained as the data set increases, we have also performed a MC simulation (see Sect. \ref{mcmcresults}). 
From those simulations we can say that our models will be reliable for up to three years, after which we will require them to be retrained with different models and training data. 
Thus, based on our analysis, these chosen models are best suited for the data at hand.
}

\subsection{LGRBs versus SGRBs}

Contemporaneously, this research can also shed light on the distinction between various GRB classes. 
{
As mentioned previously, LGRBs and SGRBs are characterized by different properties.
Thus, one has to classify GRBs as accurately as possible {in order to conduct high-precision population studies}.}
{Traditionally, the division between collapsar and merger type (non-collapsar) origin has been discussed mainly in terms of the $T_{90}$ distribution for the BATSE observed events. 
Collapsar events are characterized by $T_{90} \geq 2s$, while non-collapsar events by $T_{90}<2s$. 
An updated division based on $T_{90}$ observed by Swift has been discussed in \cite{Bromberg2013ApJ...764..179B}, where the $T_{90}$ cut has been set to 0.8.
With the advent of gravitational waves (GW) observed with the case of GRB 170817 and the consequent observation of the kilonovae, it was discussed that the merger of neutron stars was the smoking gun event to distinguish between collapsar and non-collapsar and should be accompanied by the GW and the kilonova presence. 
However, it has also been discussed extensively how the case of GW 170817 was peculiar and an unusually low luminosity event.
Moreover, there have been two GRB cases that have properties of LGRBs in terms of $T_{90}>$2s, but there was evidence of an associated kilonovae (KN): these are the instances of GRB 211211A and GRB 230307A.
The latter case has a larger value for the luminosity of the KN=$10^{42}erg s^{-1}$.
}

\subsection{Issue of LLGRBs}

{This leads us to another pertinent question about the true nature and origin of LLGRBs. Research by }\cite{Coward2005MNRAS.360L..77C,Liang2007ApJ...662.1111L,Bernardini2007A&A...474L..13B,Virgili2009MNRAS.392...91V,patel2023MNRAS.523.4923P} suggests that LLGRBs represent a distinct class of phenomena. 
Due to their low luminosities, these events are typically observed at low redshifts, such as GRB 060218 at $z=0.033$ \citep{Mirabal2006ApJ...643L..99M} and GRB 980425 at $z =0.0085$ \citep{Tinney1998IAUC.6896....3W}.
\cite{Coward2005MNRAS.360L..77C} simulated the local density rate of LLGRBs to be 220 Gpc$^{-3}$yr$^{-1}$, based on the under-luminous GRB 980425. Under the assumption that the luminosity of GRB 980425 was typical of such a population of LLGRBs, the authors claim the luminosity cutoff between LLGRBs and high-luminosity GRBs (HLGRBs) to be $2\times10^{-53}$erg s$^{-1}$.
However, based on GRB 980425 and GRB 060218 \cite{Liang2007ApJ...662.1111L} defined LLGRBs as GRBs with luminosity $<$ 10$^{-49}$erg s$^{-1}$.
The authors found the LLGRB event rate to be $\sim325^{+352}_{-177}\,\rm Gpc^{-3} yr^{-1}$ while that of HLGRBs to be $1.12^{+0.43}_{-0.2}$ Gpc$^{-3}$ yr$^{-1}$. 
{They also showed that the LFs of LLGRBs and HLGRBs are different and cannot be represented by a single function.}
\cite{Chapman2007MNRAS.382L..21C} also investigated the local density rate of LLGRBs based on GRB 980425 and GRB 060218 and found a result of $700 \pm 360$ Gpc$^{-3}$ yr$^{-1}$ within a distance of 155 Mpc.
\cite{Virgili2009MNRAS.392...91V} provided further evidence that LLGRBs are a distinct class of GRBs from HLGRBs. 
They demonstrated that a simple power-law LF is insufficient and that a smooth broken power-law fit is necessary to obtain a good prediction for the distribution of a combined sample of LLGRBs and HLGRBs.
Using MC simulation and the LF, they derive an event rate of $\sim100-400$ Gpc$^{-3}$ yr$^{-1}$ for LLGRBs and $\sim1$ Gpc$^{-3}$ yr$^{-1}$ for HLGRBs.
However, they conclude that the sample is too small and more data are needed to draw stronger conclusions.
\cite{Dainotti2024ApJ...967L..30D} obtained a GRB event rate of $8.47 - 9$ Gpc$^{-3}$ $yr^{-1}$ between $1.9<z<2.3$ using $z_{obs}$. This matches the HLGRB rate found by \cite{Dong2023ApJ...958...37D,Liang2007ApJ...662.1111L,Virgili2009MNRAS.392...91V}.
\cite{Lan2021MNRAS.508...52L} utilized 5 LLGRBs out of a collection of 1111 LGRBs obtained from Swift to demonstrate that a triple power-law LF provides a superior fit for the data, again indicating the existence of LLGRBs as a separate component.
Recently, \cite{Dong2023ApJ...958...37D} demonstrated that the event rate of HLGRB closely follows the SFR, while that of LLGRB shows an excess at low-$z$ ($z<1$). Combining the sample leads to the observed low-$z$ excess and mismatch between the GRBFR and SFR.
These previous works strongly suggest that LLGRBs are a distinct class of GRBs with the potential to solve the GRB-SFR discrepancy.
However, to draw more definitive conclusions, a larger sample of GRBs with redshift is needed.

\subsection{Discrepancy with star formation rate}

This discrepancy with the SFR could also be due to the effects of collimated emission within narrow jets not being taken into account in the current analysis.

Indeed, we need such collimated emission to achieve the observed high luminosity.
Unlike most parametric forward fitting methods, which assume both no luminosity evolution and that the GRBFR should be the same as SFR, the approach of \cite{Dainotti2024ApJ...967L..30D} offers the advantage of deriving the formation rate evolution non-parametrically directly from the data. 
Thus, we can safely bypass the assumption about the GRB rate following the SFR.
This result is also shown for LGRBs by the Konus-WIND team \citep{Tsvetkova2017ApJ...850..161T} and the other 5 teams independently \citep{Petrosian2015,Yu2015,Pescalli2016,2019MNRAS.488.5823L,yonetoku2004}, using the same statistical methods adopted by \cite{Dainotti2021ApJ...914L..40D}.
Regarding SGRBs, including those with extended emission, \cite{Dainotti2021ApJ...914L..40D} found that SGRB rates also do not follow the SFR at low-$z$, even when adding a delay for SGRBs to account for the older age of the progenitors.

\subsection{Impact of our research}

Our research here can help address the above-mentioned issues.
First, employing 276 GRBs with $z_{pred}$ will allow the community to distinguish between GRB rate and SFR and what the GRB progenitors are, specifically regarding LLGRBs.
Furthermore, the web app will enable the whole community to quickly obtain $z_{pred}$
and this will enhance the above-mentioned advantages.
The web app's versatility will help it stay updated when new GRBs with $z_{obs}$ are obtained, leading to more accurate predictions.

Finally, our MC simulations and analysis of SVOM's upcoming data paint a very promising future for GRB cosmology.
Our analysis, combined with comprehensive progenitor studies \citep{Dainotti2020ApJ...904...97D,Wang2022ApJ...924...97W}, will provide a set of sources that can be used as reliable cosmological tools.
The high-$z$ GRBs that THESEUS will explore (expected launch 2032, \cite{Theseus2021ExA....52..183A}) are crucial probes for investigating the SFR at the epoch of reionization and investigate the number of Population 3 stars.

\section{Summary and conclusion}\label{conclusion}

In this study, we have built upon the methodology established by \cite{Dainotti2024ApJ...967L..30D} and \cite{Dainotti2024GRBRedshift}.
We used a 20\% larger training set and more robust formula estimation to predict the redshifts of 276 GRBs.
In addition, we performed an extensive MC simulation to evaluate how our ML model will perform in the upcoming years.
For the benefit of the community, we have produced a publicly available and free-to-use web app designed to provide pseudo-redshifts for any GRB with plateau emission, along with other additional functionalities.
From this reasearch, we draw the following conclusions:

\begin{itemize}

    \item     The ML results obtained in this analysis, presented in Sec. \ref{mlresults}, have a similar accuracy with those presented in \cite{Dainotti2024GRBRedshift} but are slightly lower. 
    Specifically, we have a 10\% decrease in $r$, from 0.719 to 0.644, and a 9.8\% increase in RMSE, from 0.91 to 1.
    These changes in the metrics can be attributed to two reasons. 
    First, the new sample of GRBs added to the training set has a higher percentage of GRBs at a larger redshift, as compared to the training set in \cite{Dainotti2024GRBRedshift} (see Sec. \ref{datasample}). 
    This can lead to the ML model having to adjust to the new data with a larger intrinsic scatter, thus leading to a decrease in the performance metrics. 
    The second reason is the change in the methodology for formula estimation. 
    In \cite{Dainotti2024GRBRedshift}, the formulas estimated for GAM and GLM were determined using a single training-test split instance. 
    Here, we expanded the methodology and generalized the formulas. 
    Generalizing the formula here means it can better capture the underlying relations among GRB properties for a larger population of GRBs rather than just a small, limited sample of the training set.
    However, this generalization leads to poorer performance on a singular training-test split, as the variability of the underlying relation is higher in a smaller sample.
    Nonetheless, the formulas generalize well to additional synthetic GRB samples with a more uniform redshift distribution, which is evident from the improved balanced sampling result across all the metrics (lower-right and -left panels of Fig. \ref{fig:superlearner_results}).
    This promising result indicates that our ML model has effectively generalized to the entire GRB sample, and the MC simulations further emphasize this point.

    \item 
   In Section \ref{relimp_result}, we demonstrated that $\log(NH)$ exerts the greatest influence on redshift predictions. This finding is consistent with the results reported in \cite{Dainotti2024GRBRedshift} and \cite{Dainotti2024ApJ...967L..30D}.
   A similar result was also discovered independently by \cite{Racz2023CoSka..53d.100R}, who found that the performance of their ML model was reduced when they removed $\log(NH)$ from their list of predictors.
   We also find that $\log(T_a)$ is the second most important predictor of redshift, similar to \cite{Dainotti2024GRBRedshift}. However, we note that \cite{Dainotti2024ApJ...967L..30D} found that $\log(F_a)$ is the second most predictive feature. Given the correlation of $\log(T_a)$ and $\log(F_a)$, these results demonstrate that plateau features are important when predicting the redshift of GRBs.
    
    \item {In Sec \ref{generalization_results}, we predicted the redshift of 276 GRBs using our trained model. Both the bias-corrected and non-bias-corrected predictions are freely available to the community through this publication. 
    However, we still have a discrepancy between the predicted redshift distributions and the redshift distribution of the training set.}
    \cite{Dainotti2024GRBRedshift} questioned the extent the variation between the redshift distributions of GRBs with and without $z_{obs}$ are inherently different and whether the variation might be attributed to the limited size of the training set.
    {In this analysis, we have increased the training set with 30 additional GRBs. 
    However, we still need to investigate the origin of the different distributions.
    A KS test comparing the $\log(NH)$ distributions for the training and the generalization sets still presents us with the conclusion that they are derived from different parent populations. 
    Given the fact that $\log(NH)$ is the most important predictor, this could explain the discrepancy in the redshift distributions.}

    \item {The MC simulations of future data in Sec. \ref{mcmcresults} present us with promising results. 
    Based on those estimates, we conclude that our ML model is robust enough to last for at least three more years before major retraining is needed.
    The histogram distributions for $r$, RMSE, and MAD show that our ML model performs better when more sources are added to the training set. 
    It is important to note that this condition holds only if the additional data incorporated into the training set in the future comes from the same parent population as the existing sample.}
 
    \item The methodology described here can be accessed via the associated web app. 
    The app enables users to obtain redshift predictions based on the ML model presented here for their GRBs. 
    It also enables users to choose from a set of predefined models that, in our experience, work well with GRB samples and to subsequently train the models with their data.
    Then, they can obtain redshift predictions from the customized model built using the web app.
    The web app is designed to provide results similar to those presented in this work for any custom data the user may input.

    \item In Sec \ref{future}, we presented calculations on how many GRB's pseudo-redshifts we will be able to obtain in the future using SVOM data and possibly THESEUS.
    The numbers mentioned there are based on current estimates from publications of SVOM and Swift's observational statistics. 
    These numbers will need to be updated as more GRBs by SVOM are detected.
    Regardless, the numbers paint a promising picture of the future applicability of such ML techniques. 
    In fact, this work has already enabled us to increase the Swift GRB sample with redshift by roughly 110\%.

\end{itemize}

 In conclusion, we envision our web app and ML based redshift estimates will help the community in follow-up new GRBs in the coming years and the open source nature of the web app will help foster continuous development and allow new discoveries.

\section{Data availability}

{The web app is available for free on the following Github page:\url{https://github.com/gammarayapp/GRB-Web-App}.}

\begin{acknowledgements}\label{sec:acknowledgments}
The authors thank Omar Elsherif for his contributions to building the basic structure of the web app, upon which the whole functionality is based.
We gratefully acknowledge Polish high-performance computing infrastructure PLGrid (HPC Center: ACK Cyfronet AGH) for providing computer facilities and support within computational grant no. PLG/2024/017205.
We acknowledge Biagio de Simone and Christopher Cook for helping with the MC computations.
We acknowledge funding from Swift GI Cycle XIX, grant number 22-SWIFT22-0032.
Numerical computations were in part carried out on Small Parallel Computers at the Center for Computational Astrophysics, National Astronomical Observatory of Japan.

\end{acknowledgements}


\bibliographystyle{aa} % style aa.bst
\bibliography{bibliography} % your references Yourfile.bib

\begin{appendix}\label{sec:appendix}

\section{Distribution of training sets}

\begin{figure}
    \centering
    \includegraphics[width=0.89\linewidth]{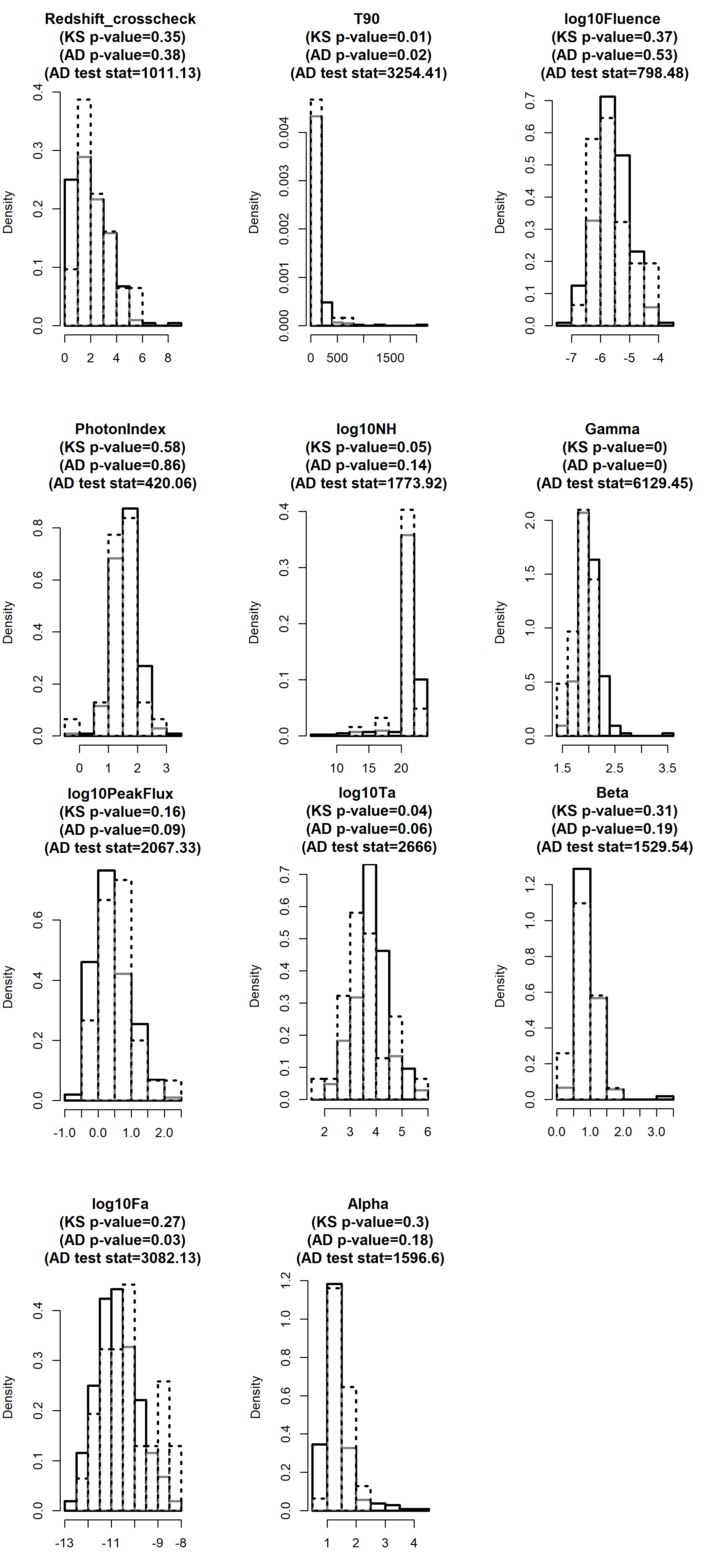}
    \caption{\textit{Distribution comparing the current training set with the previous training set from \cite{Dainotti2024GRBRedshift}.
    At the top of each distribution, the KS test and AD test comparison results are shown.}
    }
    \label{fig:datacomparision}
\end{figure}

Here, we are presenting the histogram distributions of the training set from \cite{Dainotti2024GRBRedshift} and the training set used in this work. 
The Fig. \ref{fig:datacomparision} shows the previous sample with solid lines and the new sample with dashed lines.
Each feature has the KS test and AD test statistics mentioned at the top of the histogram.
As mentioned in Sec \ref{datasample}, $T_{90}$, and $\gamma$ obtain p-values $<$ 0.05 in both AD test and KS test, while $\log(F_a)$ obtains p-value $<$ 0.05 only in AD test.
The case  can be explained by looking at the distribution, where we see a heavy-tailed distribution for the newer data set (dashed lines). Since the AD test is more sensitive to heavy-tailed distributions, we obtain a p-value of $<$ 0.05.

\FloatBarrier

\section{Weight distribution by SuperLearner}

\begin{figure}
    \centering
    \includegraphics[width=\linewidth]{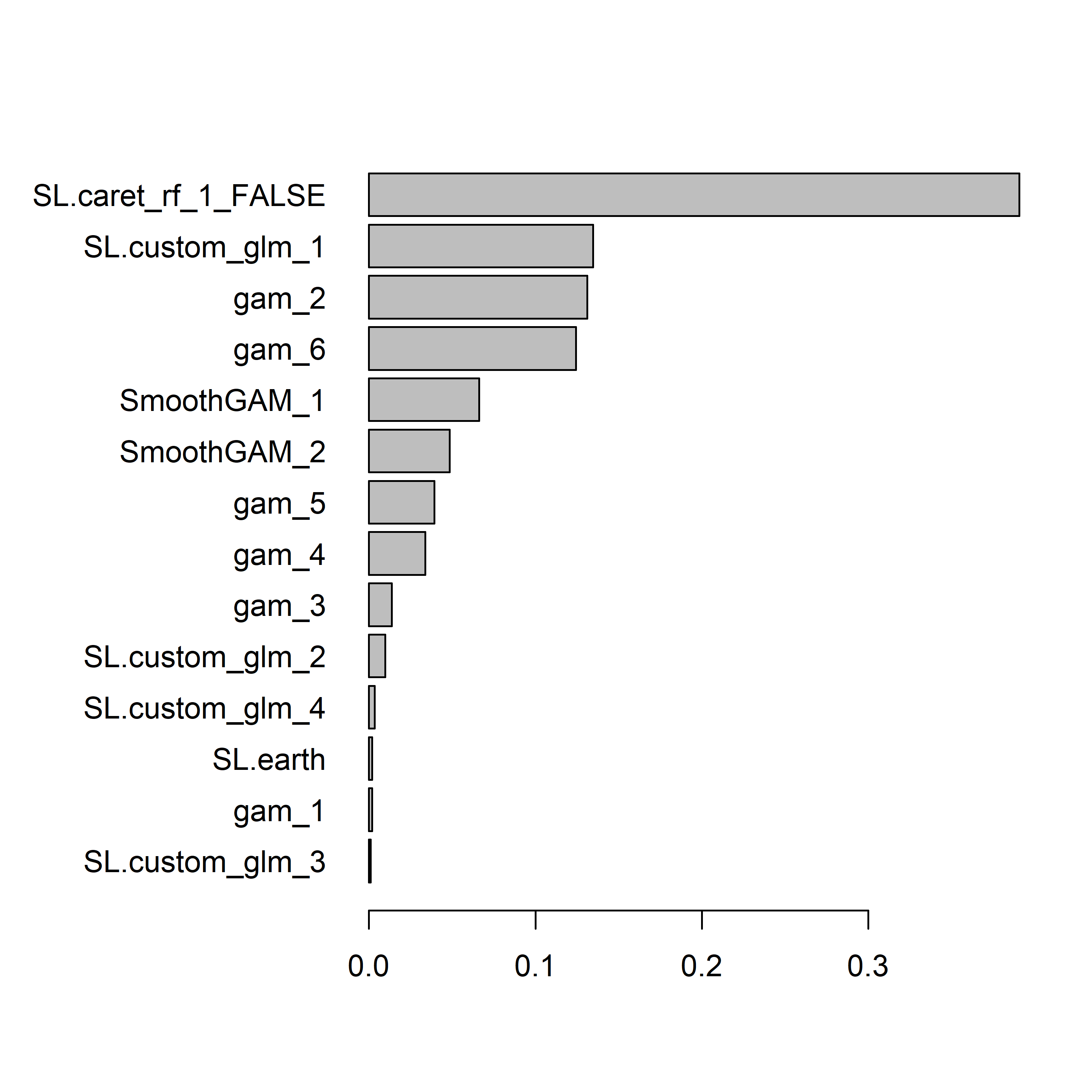}
    \caption{\textit{Weights assigned by SuperLearner to an ensemble of GAM, GLM, RF, and EARTH models.}}
    \label{model_weight}
\end{figure}

The weights assigned by SuperLearner to its constituent models vary depending on the data set it is being used.
Fig.\ref{model_weight} shows the weight assigned to EARTH is lower than those of RF, GAM, and GLM models.
However, in \cite{Narendra2022}, EARTH was the highest-weighted model.
Thus, we can say that our final ensemble model is well-suited for the size of our data sample.

\end{appendix}

\end{document}